\documentclass[preprint,showpacs,preprintnumbers,amsmath,amssymb,superscriptaddress]{revtex4}
\usepackage{graphicx,color}
\usepackage{axodraw}
\usepackage{amsmath,amssymb}
\usepackage{url}
\usepackage{epstopdf}
\newcommand{\hs}{\hspace*{0.5cm}}

\newcommand{\be}{\begin{equation}}
\newcommand{\ee}{\end{equation}}
\newcommand{\bea}{\begin{eqnarray}}
\newcommand{\eea}{\end{eqnarray}}
\newcommand{\nn}{\nonumber}
\newcommand{\crn}{\nonumber \\}

\newcommand{\al}{\alpha}
\newcommand{\la}{\lambda}
\newcommand{\bet}{\beta}
\newcommand{\ga}{\gamma}

\newcommand{\om}{\omega}
\newcommand{\pa}{\partial}
\newcommand{\fr}{\frac}

\newcommand{\bc}{\begin{center}}
\newcommand{\ec}{\end{center}}
\newcommand{\Ga}{\Gamma}

\newcommand{\ep}{\epsilon}

\newcommand{\ph}{\phi}

\newcommand {\ba}{\begin{array}}
\newcommand {\ea}{\end{array}}
\newcommand{\ben}{\begin{enumerate}}
\newcommand{\een}{\end{enumerate}}
\usepackage{bm}
\usepackage{dcolumn}

\begin{document}

\title{Electroweak theory based on $ \mbox{SU}(4)_L \otimes \mbox{U}(1)_X$ gauge group }

\author{H. N. Long}\email{hoangngoclong@tdt.edu.vn}
\address{Theoretical Particle Physics and Cosmology Research Group, Ton Duc Thang University, Ho Chi Minh City 700000, Vietnam}
\address{Faculty of Applied Sciences,
 Ton Duc Thang University, Ho Chi Minh City 700000, Vietnam}
 %%%%
\author{L. T. Hue}
\email{lthue@iop.vast.vn}
\affiliation{Institute of Research and Development, Duy Tan University,
   Da Nang City 550000, Vietnam}
 %% %
\affiliation{Institute of Physics,   Vietnam Academy of Science and Technology, 10 Dao Tan, Ba
Dinh, Hanoi 100000, Vietnam }
%--
\author{D. V. Loi}\email{dvloi@grad.iop.vast.vn}
\affiliation{Institute of Physics,   Vietnam Academy of Science and Technology, 10 Dao Tan, Ba
Dinh, Hanoi 100000, Vietnam }
\affiliation{Faculty of Mathematics-Physics-Informatics, Tay Bac University, Quyet Tam, Son La 360000, Vietnam}
\begin{abstract}
This paper includes
two main parts. In the first part, we present  generalized
 gauge models based on the  $SU(3)_C \otimes  SU(4)_L \otimes U(1)_X$ (3-4-1) gauge group with arbitrary electric charges of exotic leptons. The  mixing matrix of neutral gauge bosons is analyzed, and the eigenmasses and eigenstates are obtained.  The anomaly-free  as well as matching conditions are  discussed precisely. In the second part, we present a  new development of the original  3-4-1 model \cite{flt,pp}.
   Different from previous  works, in this paper the neutrinos, with the help of the scalar decuplet  $H$, get the   Dirac masses at the tree level.  The  vacuum expectation  value (VEV) of the Higgs boson field   in the decuplet $H$  acquiring the VEV
   responsible  for neutrino Dirac mass  leads  to mixing in  separated pairs of singly charged  gauge bosons,  namely  the  Standard Model(SM) $W$  boson and $K$, the   new gauge boson acting in the right-handed lepton sector, as well as  the singly charged bileptons $X$ and $Y$.
Due to the mixing, there occurs a right-handed current carried by the $W$ boson. From the expression of the electromagnetic coupling constant, ones get
 the limit of the sine-squared of the Weinberg angle, $\sin^2 \theta_W < 0.25$ and a constraint on electric charges of extra leptons.
 In the limit of lepton number conservation, the Higgs sector contains all massless Goldstone bosons for massive gauge bosons
and the SM-like Higgs boson.
Some phenomenology is discussed.

\end{abstract}
\maketitle
%%%%%%%%%%%%%%%%%%%
\section{Introduction}
\label{sec:intro}

The current status of particle physics leads to widespread evidence for  extending the SM.
The recently observed 750 GeV diphoton excess \cite{diphoton} can be explained  as an existence of
a new neutral scalar   that couples  to extra heavy quarks, or in some cases, to new leptons and bosons.
In this sense, the 3-3-1 models \cite{331ppf,flt,331flt} seem to be  good candidates since they contain all ingredients
such as extra quarks and new scalar fields.
However,  the problem is that satisfying the LHC  diphoton excess and the experimental value of the muon anomalous magnetic moment $(g-2)_\mu$ requires that the 3-3-1 scale  be
 low $\om \approx 400$ GeV, while the flavor-changing neutral current (FCNC) requires a
high scale $\om \approx 2$ TeV.    To solve this
puzzle for the 3-3-1 model with right-handed neutrinos, one must introduce an inert  scalar triplet \cite{boucena},  extend the gauge
 group to a larger one such as  $SU(3)_C \otimes SU(3)_L \otimes SU(3)_R \otimes U(1)_X$
 \cite{huongdong750}, or  introduce  new  charged scalars \cite{tq}.

The above situation is also correct for the  other SM  extensions; hence the search for the models
 satisfying current experimental data is needed.
It is known that the $SU(4)$ is the highest symmetry group in the electroweak sector \cite{o341}. There have been some
 gauge models based on the $SU(3)_C \otimes  SU(4)_L \otimes U(1)_X$ \cite{flt,pp,palcu}; however,
 the Higgs physics - currently  the  most important sector,  has not received enough attention. In some versions,  this puzzle was not studied in much detail at all. In light of  the  current  status of particle physics, the Higgs sector should be considered with as much detail as possible, especially in
the neutral scalar sector where  the SM-like Higgs boson is contained. Thus in this work, we will focus on the Higgs sector. As in the 3-3-1 models, the electric charge quantization was also  explained  in
 the framework of the 3-4-1 model \cite{eq}.

The aim of this paper is to present the 3-4-1 models which are able to deal with current Higgs physics.
It is well known that if scalar sector contains
 many neutral scalar fields the situation is very complicated. In addition, we will pay attention to
  the gauge boson sector where new physics is quite rich
 and explore some interesting features of phenomenology. The 3-4-1 model we consider here may be considered as
 combination of the minimal
 3-3-1 model \cite{331ppf} and an alternative version with right-handed neutrinos \cite{flt,331flt}.
 Hence, the derived model is quite interesting and deserves further study.

Our work is arranged as follows. Section \ref{model} will  present conditions of anomaly cancellation
 and particle content, where
extra novel induced leptons have arbitrary electric charges $q$ and $q^\prime$. We will show that the 3-4-1 models are anomaly free only
 if there are an equal number of  quadruplets ($\textbf{4}$) and antiquadruplets ($\textbf{4}^*$). Simply speaking,
 this condition  requires   the sum of all fermion charges to vanish (see subsection \ref{anomaly}). In subsection \ref{yukawa}, the Higgs bosons
  needed for fermion mass  production are discussed. We want to avoid nonrenormalizable effective couplings, so every Higgs
   multiplet has only one component  with nonzero VEV. Subsection \ref{gaugeboson}focus on the gauge boson fields,
especially neutral ones. Section \ref{m341rhn} is devoted to the
 original 3-4-1 model \cite{flt,pp}. As in \cite{pp}, to produce the mass for leptons, the Higgs decuplet is introduced. However,
as will be seen below, it has to be redefined. In this work, we will build a lepton number operator from which the lepton
 flavor number violating (LFV) processes will be pointed out. In this section, mixing of the singly charged gauge bosons
 and the currents will get more attention.  For completeness, the neutral gauge boson sector  will also be  presented,  though it is quite
similar to the previous analysis. From an  expression of the ratio of two gauge couplings $t$, it follows the bound on the  sine-squared of the Weinberg angle
$\sin^2 \theta_W < 0.25$.   In section \ref{phen} the bounds on masses of new gauge bosons are
roughly derived, based on the  data of $W$ bosons, rare muon decays and  $\mu-e$ conversion.
Finally, we present our conclusion in Sec. \ref{conc}.

\section{The model}
\label{model}
As above mentioned, we first check the conditions for anomaly free of the models based on
  $\textrm{SU}(3)_C \otimes  \textrm{SU}(4)_L \otimes \mbox{U}(1)_X$ gauge group.
\subsection{\label{anomaly}Anomaly cancellation and fermion content}
For the class of the models constructed from the gauge group  $\textrm{SU}(3)_C\times \textrm{SU}(3)_L\times \textrm{U}(1)_N$, the conditions for anomaly cancellation were discussed in detail in \cite{anomaly}. Similarly for the class of the  $\textrm{SU}(3)_C\times \textrm{SU}(4)_L\times \textrm{U}(1)_X$ (3-4-1) models the following gauge anomalies must vanish: i) $[SU(3)_C]^2\times U(1)_X$, ii) $[SU(4)_L]^3$, iii) $[SU(4)_L]^2\times U(1)_X$; iv) $[\mathrm{Grav}]^2\times U(1)_X$;  and v) $[U(1)_X]^3$.   Being different from the studies of anomaly cancellation for the 3-3-1 models \cite{anomaly,afree}, we will  exploit the relation between charge operator and diagonal generators of the gauge symmetry $SU(4)_L$  to prove that the five conditions will reduce to two conditions only: $[SU(4)_L]^3$ and $[SU(4)_L]^2\times U(1)_X$.

For a general 3-4-1 model, the electric charge operator is in the form
\be Q = T_3 + b T_8 + c T_{15} + X\, ,
\label{m4eq1}
\ee
where the coefficient in front of $T_3$ equaling 1, is chosen to ensure that the SM group  is a subgroup of the   model
under consideration: $SU(2)_L \otimes U(1)_Y \subset
SU(4)_L \otimes U(1)_X$.

The leptons are in quadruplet
\be f_{a L} = ( \nu_a\, , l_a \, , E_a^q \, , E_a^{\prime q'} )_L^T,\hs  a = e, \mu, \tau \, ,
\label{m4eq2}
\ee
where $q$ and $q'$ are electric charges of associated extra leptons. Applying Eq.(\ref{m4eq1}) to Eq.(\ref{m4eq2})
we obtain
\be b = \fr{-2q -1}{\sqrt{3}} \, , \hs c = \fr{q - 3q' -1}{\sqrt{6}} \, , \hs X_{f_{aL} } = \fr {q + q' -1}{4},
\label{m4eq3}
\ee
or
%--
\be
q=-\frac{1}{2}-\frac{\sqrt{3}b}{2}, \hs q'=-\frac{1}{2}-\frac{b}{2\sqrt{3}}-\frac{\sqrt{2}c}{\sqrt{3}}\hs \mathrm{and}\; X_{f_{aL}}=-\frac{1}{2}-\frac{b}{2\sqrt{3}}-\frac{c}{2\sqrt{6}}.
\label{qqp}\ee
 Before discussing anomaly cancellation, we remember that the fermion representations of $SU(3)_C$ and $SU(4)_L$ in the 3-4-1 models are all $SU(3)_C$ triplets, $SU(4)_L$ (anti) quadruplets,  and singlets.
All singlets do not contribute to anomalies so in consideration we omit them. The representative matrices
 of generators corresponding to
the $SU(3)_C$ triplets are denoted as $T^a_C$ ($a=1,2,..,8$) , and the $SU(4)_L$
(anti)quadruplets are $T^a_L(\overline{T}^a_L)$, $a=1,2,..,15$.

Now we consider the 3-4-1 model with $M$ and $N$ families of leptons and quarks, respectively.  In addition, the number of $SU(4)_L$ quadruplets of quark families is  $K$. For simplicity we assume that all left-handed leptons are in the quadruplets. The general case is derived easily. All of them respect the gauge symmetry
 $SU(3)_C\times SU(4)_L\times U(1)_X$ as follows. For leptons, we have
\bea
f_{iL} &=& (\nu_{iL},l_{iL}, E^{q}_{iL}, E'^{q^\prime}_{iL})^T\sim (1, 4, X_{f_{L}}),\crn
 \nu_{iR}&\sim& (1, 1, 0), \; l_{iR}\sim (1, 1,-1), \; E^q_{iR}\sim (1, 1,q), \;
  E'^{q^\prime}_{iR}\sim (1, 1,q^\prime),  \hs i=1,2,...,M. \label{leptonreps}
  \eea
  These left-handed leptons are generalized from (\ref{m4eq2}).  As in the 3-3-1 models \cite{flt,331flt,331ppf},
   the parameters  $b$ and $c$ are closely connected with $q$ and $q'$;   in \cite{anyqq1,anyqq},
   the parameters $b,c$ have been used.  In our point of view, the clearer way is
  using  $q$ and $q'$.  We note that  $\nu_{iR}$ is an option,
  while $E^q_{iR}$ and $E^{q'}_{iR}$ may disappear in the specific case of  the minimal 3-4-1 model.
  If the left-handed leptons are in antiquadruplet,
  then   $b$ and $c$ will be replaced by $-b$ and $-c$, respectively.

  The quark sector is
  \bea Q_{mL}&=&(u_{mL},d_{mL}, T_{mL}, T'_{mL})^T \sim (3, 4, X_{q_{L}}),\hs \; m=1,2,...,K,\crn
     Q_{nL}&=&(d_{nL}, -u_{nL}, D_{nL}, D'_{nL})^T \sim (3, 4^*,X_{\bar{q}_{L}}),\crn
     u_{pR}&\sim&  (3, 1, 2/3), \; d_{pR}\sim  (3, 1, -1/3), \; T_{mR}\sim  (3, 1, X_{T_R}),\;
     T'_{mR}\sim  (3, 1, X_{T'_R}),\crn
   D_{nR}&\sim&  (3, 1, X_{D_R}),\; D'_{nR}\sim  (3, 1, X_{D'_R}),\; n=K+1,.., N, \; p=1,2,.., N.
   \label{quarkrep}\eea

Let us first consider the anomaly of $[SU(4)_L]^3$. Each of the $SU(4)_L$ quadruplets $4_L$ or
 antiquadruplets $4^*_L$ contributes a well-known quantity $\mathcal{A}^{abc}(4_L)=
 \mathrm{Tr} (T^a_L\{T^b_L,T^c_L\})$ or $\mathcal{A}^{abc}(4^*_L)=\mathrm{Tr} (\bar{T}^a_L
 \{\bar{T}^b_L,\bar{T}^c_L\})$, where $a,b,$ and $c$ mean three $SU(4)_L$ gauge
 bosons related to the triangle diagrams. Because $\mathcal{A}^{abc}(4_L)=
 -\mathcal{A}^{abc}(4^*_L)$, the total contribution to the $[SU(4)_L]^3$ anomaly can be written as
\be \mathcal{A}^{abc}(4_L)\left( \sum_{Q_{mL},f_{iL}}4_L-\sum_{Q_{nL}}4^*_L\right)=
 \mathcal{A}^{abc}(4_L)\left( n_{4_L}-n_{4^*_L}\right),\label{su43}  \ee
where $n_{4_L}$  and $n_{4^*_L}$ are  the number of fermion quadruplets
and antiquadruplets, respectively. This means that the above anomaly cancels only if the number of
quadruplets is equal to the number of antiquadruplets, namely
\be M+6 K=3N \label{su4la},\ee
where the factor 3 appears because the quarks are in  $SU(3)_C$ triplets while leptons are
in $SU(3)_C$ singlets.

Next, we consider the anomaly of $[SU(4)_L]^2\times U(1)_X$. The  $[SU(4)_L]^2$ gives the
same factor for both quadruplets and anti - quadruplets, i.e., $\mathrm{Tr}[T^aT^b]=
\mathrm{Tr}[\bar{T}^a\bar{T}^b]=\delta_{ab}/2$, where $a$ and $b$ relate to two
$SU(4)_L$ gauge bosons.  Hence, this anomaly is free if the sum of all $U(1)_X$ charges
of the $SU(4)_L$ chiral multiplets is zero, namely,
\be \sum_{f_{iL},Q_{pL}}X_{L}= M X_{f_L}+ 3K X_{q_L} +3(N-K) X_{\bar{q}_L}=0, \label{su42u11a}\ee
where $X_{L}$  denotes the $U(1)_X$ charge of an arbitrary left-handed (anti) - quadruplet in the model.

Now we turn to the anomaly of the $[SU(3)_C]^2\times U(1)_X$. This case is similar to the case of
$[SU(4)_L]^2\times U(1)_X$, but now only the $SU(3)_C$ quark triplets contribute to the mentioned anomaly.
The anomaly-free condition is
\be  \sum_{Q_{mL}}4X_{q_L} + \sum_{Q_{nL}}4X_{\bar{q}_L}- \sum_{q_R}X_{q_R}=0,
\label{SU3c2u1a}\ee
where $X_{q_R}$  and   $X_{q_L,\bar{q}_L}$ are  $U(1)_X$ charges of
  right-handed $SU(4)_L$ quark singlets $q_R$ and left-handed $SU(4)_L$ quark (anti)quadruplets $Q_{m_L}(Q_{n_L})$, respectively.
The factors 4 appear in Eq. (\ref{SU3c2u1a}) because we take four components of every $SU(4)_L$ (anti)quadruplet  into account.
  The minus sign implies the opposite contributions of left- and right-handed
 fermions to gauge anomalies. Because all $q_{R}$ are singlets of the $SU(4)_L\times U(1)_X$,
  their $U(1)_X$ charges are always equal to the electric charges $q_{q_R}$, leading
  to $\sum_{q_R}X_{q_R}=\sum_{q_R}q_{q_R}$. On the other hand, from the definition
  of the charge operators $Q$  given in (\ref{m4eq1}), it can be seen that
  \[ \sum_{Q_{mL}}4X_{q_L}=  \sum_{Q_{mL}}\mathrm{Tr}\left(X_L\times I_4\right)
  =  \sum_{Q_{mL}} \mathrm{Tr}[Q ],\]
  where $\mathrm{Tr}[Q ]$ implies the sum over  the electric charges of  all components of
   the quark quadruplet  $Q_{mL}$.  Note that we have used the traceless property
   of the $SU(4)_L$ generators: $\mathrm{Tr}(T^a)=0$.  Doing this the same way for the
   case of quark antiquadruplets, the condition (\ref{SU3c2u1a}) can be rewritten in
   terms of the electric charges of  left-and right-handed quarks,
  \be   \sum_{Q_{pL}}\sum_{i=1}^4q_{ q_{L}}- \sum_{q_R}q_{q_R}=0,\label{su42u11ap}\ee
  where the first sum implies that all $SU(4)_L$ quark (anti)quadruplets and
  their components are counted. The equality (\ref{su42u11ap}) is always correct because
   every left-handed quark  always has its right-handed partner with the same electric charge.

The above discussion on the anomaly cancellation of $[SU(3)_C]^2\times U(1)_X$ can be
applied for the case of the anomaly cancellation of  $[\mathrm{Grav}]^2\times U(1)_X$,
but the electric charges must be counted for all components of the $SU(3)_C$ and $SU(4)_L$
multiplets of all quarks and leptons. The proof  of the zero contribution of  the quark sector
is very  easy while that of the lepton sector needs more explanation.  Although the
 presence of right-handed neutrinos is optional, they are neutral leptons and, therefore,  do not contribute
  to this anomaly.  If right-handed charged leptons are arranged into components of left-handed
   (anti)quadruplets, they must be changed into their charge  conjugations. These new forms
   of right-handed leptons have the opposite signs of electric charges compared to
   the respective left-handed  partners. Hence, the total contribution to the considered  anomaly is still zero.

Cancellation of  the $[U(1)_X]^3$ anomaly, which relates to the triangle diagram having three $B''$ gauge bosons, is expressed  by the following condition:
  \be \sum_{F_L} X^3_{F_L}-\sum_{F_R} X^3_{F_R}=0, \label{3u12}\ee
   where $F_L$ and $F_R$ are any components of the fermion (quark and lepton) representations of the $SU(3)_C$ and $SU(4)_L$ gauge symmetries.
   Hence, the sum is taken over all components of these  representations. Because the $U(1)_X$
   and the electric charges relate to each other through the definition of the charge operator (\ref{m4eq1}),
   we can write the left-hand side of (\ref{3u12}) as a function of electric charges.  Because all
    right-handed fermions  are $SU(4)_L$ singlets,  $X_{F_R}=q_{F_R}$; therefore,
   \be\sum_{F_R} X^3_{F_R}=\sum_{F_R} q^3_{F_R}.  \label{RHcon}\ee
   %-
     In contrast, all left-handed fermions are (anti)quadruplets, and we can write the left-handed term in  (\ref{3u12}) as a sum over all fermion (anti)quadruplets, namely,
  %---
  \be \sum_{F_L} X^3_{F_L}= \sum_{4_L,4^*_L} 4 X^3_{F_L}. \label{lhcon}\ee
  Now we come back to the formula of the charge operator, where we denote $Q_{F_L}$ as the charge
  operator of the left-handed fermion (anti)quadruplets. If we denote $T^{(3,8,15)}\equiv
   T^3+bT^8+cT^{15}$,  the $U(1)_X$ charge part of each $4_L$ representation can be written as
  \bea X_{F_L} I_4&=& Q_{F_L}- T^{(3,8,15)}   \rightarrow  \left(X_{F_L} I_4\right)^3=
  \left( Q_{F_L}- T^{(3,8,15)}\right)^3\crn
  \rightarrow \mathrm{Tr}\left[ X^3_{F_L} I_4\right]&=&   \mathrm{Tr}\left[Q^3_{F_L}\right]
  -3 \mathrm{Tr}\left[Q_{F_L}T^{(3,8,15)}(Q_{F_L}-T^{(3,8,15)})\right]-
  \mathrm{Tr}\left[\left(T^{(3,8,15)}\right)^3\right]\crn
  \rightarrow 4 X^3_{F_L} &=&   \mathrm{Tr}\left[Q^3_{F_L}\right] -
  3\mathrm{Tr}\left[(T^{(3,8,15)}+X_{F_L}I_4)T^{(3,8,15)}X_{F_L}I_4\right]
  -\mathrm{Tr}\left[\left(T^{(3,8,15)}\right)^3\right]\crn
   &=&  \mathrm{Tr}\left[Q^3_{F_L}\right] -3 X_{F_L}\mathrm{Tr}\left[(T^{(3,8,15)})^2\right]
   - \mathrm{Tr}\left[\left(T^{(3,8,15)}\right)^3\right],\label{xfl1}\eea
   where we have used the fact that  both $Q$ and $T^{(3,8,15)}$ are diagonal so they
   commute with each other, and $T^{(3,8,15)}$ is traceless. Remember that $4 X^3_{F_L}$ is the
   contribution of four components in one quadruplet.  Then the contribution to $[U(1)_X]^3$ of
   all quadruplets is
   \bea \sum_{4_L}X^3_{F_L}= \sum_{F_L} q^3_{F_L} - 3\sum_{4_L}X_{F_L}\left[(T^{(3,8,15)})^2\right]
   - n_{4_L}\mathrm{Tr}\left[\left(T^{(3,8,15)}\right)^3\right]. \label{cqurtet}\eea
 The same proof can be applied for the case of antiquadruplets with generators
 $\overline{T}^a=-T^a$, ($a=3,8,15$) and $\overline{T}^{(3,8,15)}=-T^{(3,8,15)}$.
 From this, it can be proved that
 \[\mathrm{Tr}\left[\left(\overline{T}^{(3,8,15)}\right)^2\right]=\mathrm{Tr}\left[\left(T^{(3,8,15)}\right)^2\right],\hs \mathrm{Tr}\left[\left(\overline{T}^{(3,8,15)}\right)^3\right]=-\mathrm{Tr}\left[\left(T^{(3,8,15)}\right)^3\right].\]
 %--
 The above discussion is enough to write (\ref{3u12})   in the following new form:
 \be  \left( \sum_{F_L}q^3_{F_L}-\sum_{F_R}q^3_{F_R}\right)   - 3\mathrm{Tr}\left[(T^{(3,8,15)})^2\right]
 \sum_{4_L,4^*_L}X_{F_L}- \mathrm{Tr}\left[\left(T^{(3,8,15)}\right)^3\right]
 \left( n_{4_L}- n_{4^*_L}\right) =0. \label{u13p}\ee
  The equality (\ref{u13p}) is satisfied as a consequence of the two anomaly-free
  conditions (\ref{su43}) and (\ref{su42u11a}).  The equality (\ref{su42u11a}) also implies
  that the sum over the electric charges of left-handed fermions is zero.

Finally,  we conclude  that the anomaly-free conditions of the 3-4-1 models are as follows: (i) the number
of  fermion quadruplets is  equal to that of fermion antiquadruplets, (ii)
sum over electric charges of all left-handed fermions is zero.

The 3-4-1 models, of concern here are all satisfied with these two conditions.

\subsection{\label{yukawa}Yukawa couplings and masses for fermions }
Since the leptons are arranged as
\bea f_{a L}  & = &  ( \nu_a\, , l_a \, , E_a^q \, , E_a^{\prime  q'} )_L^T \sim \left(1, 4, \fr 1 4 (q + q' -1) \right)\, ,\crn
l_{a R} & \sim &  (1, 1, -1) \, , \hs E_{a R} ^q   \sim  (1, 1, q) \, , \hs E_{a R} ^{\prime  q'} \sim (1, 1, q')\, ,
\label{m4eq6}
\eea
the mass of $E_{a } ^{\prime  q'} $ is obtained from the Yukawa coupling,
\be - L_{\mathrm{Yukawa}}^{E'} =  h^{E'}_{a b} \overline{f_{a L} } \Phi_1 E_{b R} ^{\prime  q'}  +\mathrm{ H. c.} \, ,
\label{m4eq7}
\ee
where
\be \Phi_1 \sim \left( 1, 4, \fr{(q -3 q' -1) }{4}\right) = \left(%
%\begin{array}{c}
\Phi_1^{(-q')} \, ,
\Phi_1^{(-q' -1)} \, ,
\Phi_1^{(q-q')} \, ,
\Phi_1^{0}
%\end{array}\,
\right)^T.
\label{m4eq8}
\ee
Hence, if $\Phi_1^0$ has a VEV $\fr{V}{\sqrt{2}}$, then $E_{a } ^{\prime  q'} $ gets mass from a mass matrix
\be (m_{E' })_{a b} = h^{E'}_{a b} \fr{V}{\sqrt{2}}\, .
\label{m4eq9}
\ee

The mass of $E_{a } ^{ q} $ is obtained from the following Yukawa term,
\be - L_{\mathrm{Yukawa}}^{E} =  h^{E}_{a b} \overline{f_{a L} } \Phi_2 E_{b R} ^{ q}  +\mathrm{ H. c}. \, ,
\label{m4eq10}
\ee
where
\be \Phi_2 \sim \left( 1, 4, -\fr{(1 + 3q - q' ) }{4}\right) = \left(%
%\begin{array}{c}
\Phi_2^{(-q)}  \, ,
\Phi_2^{(-q -1)} \, ,
\Phi_2^0 \, ,
\Phi_2^{(q' - q)}
%\end{array}\,
\right)^T\, .
\label{m4eq11}
\ee
Thus,  if $\Phi_2^0$ has a VEV $\fr{\om}{\sqrt{2}}$, then $E_{a } ^{ q} $ gets mass from a matrix:
\be (m_{E })_{a b} = h^E_{a b} \fr{\om}{\sqrt{2}}.
\label{m4eq12}
\ee
Finally   the ordinary lepton masses come from the following  Yukawa term,
\be - L_{\mathrm{Yukawa}}^{l} =  h^{l}_{a b} \overline{f_{a L} } \Phi_3 l_{b R}   +\mathrm{ H. c.} \, ,
\label{m4eq13}
\ee
where
\be \Phi_3 \sim \left( 1, 4, \fr{(3 + q + q' ) }{4}\right) = \left(%
%\begin{array}{c}
\Phi_3^{(+)}  \, ,
\Phi_3^{0} \, ,
\Phi_3^{(q+1)} \, ,
\Phi_3^{(q'+1 )}
%\end{array}\,
\right)^T.
\label{m4eq14}
\ee
If $\Phi_3^0$ has a VEV $\fr{v}{\sqrt{2}}$, then the mass matrix related to masses of $l_{a }$ is
\be (m_{l })_{a b}= h^l_{a b} \fr{v}{\sqrt{2}}\, .
\label{m4eq15}
\ee
%-
We turn now to the quark sector  where
\bea Q_{3 L} &  = & \left(%
%\begin{array}{c}
u_3  \, ,
d_3 \, ,
T \, ,
T^\prime
%\end{array}\,
\right)^T_L \sim \left(3, 4,    \fr{5+ 3 (q+q')}{12} \right)\, , \crn
u_{3 R} & \sim & (3, 1, 2/3 )\, , \hs  d_{3 R}  \sim  (3, 1, - 1/3 )\, , \crn
T_R & \sim &  \left(3, 1, \fr{2+3q}{3}\right) \, , T^\prime _R \sim \left(3, 1, \fr{2+3q'}{3}\right).
\label{m4eq16}
\eea
The $u_3$ gets mass through the Yukawa part,
\be - L_{\mathrm{Yukawa}}^{t} =  h^{t}\overline{Q_{3 L} } \Phi_4 u_{3 R}   + \mathrm{H. c.} \, ,
\label{m4eq17}
\ee
where
\be \Phi_4 \sim \left( 1, 4, \fr{( q + q' -1 ) }{4}\right) = \left(%
%\begin{array}{c}
\Phi_4^{0}  \, ,
\Phi_4^- \, ,
\Phi_4^{(q)} \, ,
\Phi_4^{(q' )}
%\end{array}\,
\right)^T\, .
\label{m4eq18}
\ee
If  $\Phi_4^0$ has a VEV $\fr{u}{\sqrt{2}}$, then the mass term of $u_{3 } $ is
\be m_{u_3} = h^t\fr{u}{\sqrt{2}}.
\label{m4eq19}
\ee
The other Yukawa terms related to  $Q_{3L}$ are
\be - L_{\mathrm{Yukawa}}^{g3}  = h^{b}\overline{Q_{3 L} } \Phi_3 d_{3 R} + h^{T}\overline{Q_{3 L} } \Phi_2 T_{ R} +
h^{T^\prime}\overline{Q_{3 L} } \Phi_1 T^\prime_{ R} + \mathrm{H.c.},
\label{m4eq20}
\ee
which give three  mass terms:
\be m_{d_3} = h^b \fr{v}{\sqrt{2}}\, , m_{T} = h^T \fr{\om}{\sqrt{2}}\, , m_{T^\prime} = h^{T^\prime} \fr{V}{\sqrt{2}}\, .
\label{m4eq21}
\ee
%--
Two other quark generations are
%-----
\bea Q_{\al L} &  = & \left(%
%\begin{array}{c}
d_\al  \, ,
- u_\al \, ,
D_\al \, ,
D_\al^\prime
%\end{array}\,
\right)^T_L \sim \left(3, 4^*,   - \fr{1 + 3 (q+q')}{12} \right)\, , \hs \al = 1, 2 , \crn
u_{\al R} & \sim & (3, 1, 2/3 )\, , \hs  d_{\al R}  \sim  (3, 1, - 1/3 )\, , \crn
D_{\al R} & \sim &  \left(3, 1, -\fr{1+3q}{3}\right) \, , D_{\al R}^\prime  \sim \left(3, 1, -\fr{1+3q'}{3}\right).
\label{m4eq16a}
\eea
The relevant  Yukawa terms are
\bea - L_{\mathrm{Yukawa}}^{12} & = &  h^{d2}_{\al \bet}\overline{Q_{\al  L} } \Phi_4^\dag d_{\bet R}   +
h^{u2}_{\al \bet}\overline{Q_{\al  L} } \Phi_3^\dag u_{\bet R}   + \crn
&+&  h^{D2}_{\al \bet}\overline{Q_{\al  L} } \Phi_2^\dag D_{\bet R}   +
h^{D'2}_{\al \bet}\overline{Q_{\al  L} } \Phi_1^\dag D^\prime_{\bet R} +
 \mathrm{H. c.} \, ,
\label{m4eq17a}
\eea
from which it follows that
\be (m_{d_2})_{ \al \bet} = h^{d2}_{\al \bet} \fr{u}{\sqrt{2}}\, ,  (m_{u_2})_{ \al \bet}  = -
 h^{u2}_{\al \bet} \fr{v}{\sqrt{2}}\, ,   (m_{D_2})_{ \al \bet } = h^{D2}_{\al \bet} \fr{\om}{\sqrt{2}}\, ,
   (m_{D_2^\prime })_{\al \bet } = h^{D'2}_{\al \bet}\ \fr{V}{\sqrt{2}}\, .
\label{m4eq18a}
\ee
We emphasize that if all fermions except neutrinos have the right-handed counterparts, then only
four Higgs quadruplets are needed.  Because the sum of the $X$-charges  over four Higgs quadruplets vanishes,  in the Higgs potential,
there always exists an antisymmetric  term $\ep_{ijkl} \Phi_1^i \Phi_2^j \Phi_3^k \Phi_4^l$.

\subsection{\label{gaugeboson}Gauge boson masses}

Gauge boson masses arise from the covariant kinetic term of the Higgs bosons,
\be L_{\mathrm{Higgs}} = \sum_{i=1}^4 \left(D^\mu\langle \Phi_i \rangle\right)^\dag D_\mu \langle \Phi_i \rangle \, .
\label{m4eq22}
\ee
The covariant derivative is defined as
\bea
D_\mu & = & \pa_\mu - i g\sum_{a=1}^{15} A_{a \mu} T_a  - i g' X B''_\mu T_{16} \, \crn
& \equiv & \pa_\mu - i g P_\mu^{NC} - i g P_\mu^{CC}\, ,
\label{m4eq23}
\eea
where $ g, g'$ and $ A_{a \mu} , B''_\mu$ are gauge couplings  and fields of  the gauge groups $SU(4)_L$ and $U(1)_X$, respectively.
For the quadruplet, $T_{16} = \fr{1}{2\sqrt{2}} \textrm{diag} (1,1,1,1)$, and the part relating to neutral currents is
\bea P_\mu^{NC} & = &\fr 1 2  \textrm{diag} \left(A_3 + \fr{A_8}{\sqrt{3}} + \fr{A_{15}}{\sqrt{6}} + X t  \fr{B''}{\sqrt{2}}\, , -A_3 + \fr{A_8}{\sqrt{3}} + \fr{A_{15}}{\sqrt{6}} + X t  \fr{B''}{\sqrt{2}}\right.\, ,\crn
 & & \hs \hs \hs \left.-\fr{ 2A_8}{\sqrt{3}} + \fr{A_{15}}{\sqrt{6}} + X t  \fr{B''}{\sqrt{2}}\, ,  -\fr{3A_{15}}{\sqrt{6}} + X t  \fr{B''}{\sqrt{2}}
 \right)_\mu ,
\label{m4eq24}
\eea
where the spacetime indices of gauge fields are omitted  for compactness, and $t\equiv g'/g$.
The part associated with charged currents is
\bea P_\mu^{CC} & = &\fr 1 2 \sum_{a} \la_a A_{a\mu} \, ; \hs a = 1,2,4,5,6,7,9,10,11,12,13,14\crn
&=& \fr{1}{\sqrt{2}}\left(%
\begin{array}{cccc}
0 & W^+ &W_{13}^{-q} & W_{14}^{-q'}\\
 W^-  & 0 & W_{23}^{-(1+q)}  &W_{24}^{-(1+q')}\\
W_{13}^{q} & W_{23}^{(1+q)} & 0 &W_{34}^{(q-q')} \\
W_{14}^{q'}&W_{24}^{(1+q')}&W_{34}^{-(q-q')}&0
\end{array}\,
\right)_\mu\, ,
\label{m4eq25}
\eea
where we have denoted $\sqrt{2}\, W^\mu_{13} \equiv A^\mu_1 - i A^\mu_3$ and so forth.
The upper subscripts label the electric charges of gauge bosons. We note that this part does not depend on the $X$-charges of quadruplets.

To summarize, with the following Higgs vacuum structure,
\bea  \langle \Phi_1 \rangle &= &   \left(%
%\begin{array}{c}
0 \, ,
0\, ,
0 \, ,
 \fr{V}{\sqrt{2}}
%\end{array}\,
\right)^T, \hs   \langle \Phi_2 \rangle  =    \left(%
%\begin{array}{c}
0 \, ,
0\, ,
\fr{\om}{\sqrt{2}}\, ,
0
%\end{array}\,
\right)^T, \crn
 \langle \Phi_3 \rangle &= &   \left(%
%\begin{array}{c}
0 \, ,
 \fr{v}{\sqrt{2}}\, ,
0 \, ,
 0
%\end{array}\,
\right)^T, \hs   \langle \Phi_4 \rangle  =    \left(%
%\begin{array}{c}
 \fr{u}{\sqrt{2}} \, ,
0\, ,
0\, ,
0
%\end{array}\,
\right)^T, \label{l1}
\eea
masses of non-Hermitian (charged)  gauge bosons are given by
\bea m^2_W &  = & \fr{g^2(v^2 + u^2)}{4}\, ,  m^2_{W_{13}}  =  \fr{g^2(u^2 + \om^2)}{4}\, ,
 m^2_{W_{23}}  =  \fr{g^2(v^2 + \om^2)}{4}\, , \crn
m^2_{W_{14}} &  = &    \fr{g^2(u^2 + V^2)}{4}\, , m^2_{W_{24}}  =  \fr{g^2(v^2 + V^2)}{4}\, ,
 m^2_{W_{34}}  =  \fr{ g^2(\om^2 + V^2)}{4}\, .
\label{m4eq34}
\eea
By  spontaneous symmetry breaking (SSB), the following relation should be in order:
 $V\gg \om \gg u, v$, and from (\ref{m4eq34}) one gets
 \be u^2 + v^2 = v_{SM}^2 = 246^2 \hs \textrm{GeV}^2. \ee
\subsection{Neutral gauge bosons}
Inserting Eq.(\ref{m4eq24}) into the Higgs multiplets, we get
 mass terms
\bea  M_{\mathrm{mass}}^{2NG}  & = &
  \fr{g^2}{4} \left[u^2\left(A_3 + \fr{A_8}{\sqrt{3}} + \fr{A_{15}}{\sqrt{6}} + X_{\Phi_4} t  \fr{B''}{\sqrt{2}}\right)^2\right. \crn
&+ & v^2\left(-A_3 + \fr{A_8}{\sqrt{3}} + \fr{A_{15}}{\sqrt{6}} + X_{\Phi_3} t  \fr{B''}{\sqrt{2}}\right)^2\crn
&+&\left.  \om^2 \left(- \fr{2A_8}{\sqrt{3}} + \fr{A_{15}}{\sqrt{6}} + X_{\Phi_2} t  \fr{B''}{\sqrt{2}}\right)^2
+ V^2 \left(- \fr{3 A_{15}}{\sqrt{6}} + X_{\Phi_1} t  \fr{B''}{\sqrt{2}}\right)^2
\right].
\label{m4eq36}
\eea
In the basis $\left(A_{3 \mu}, A_{8 \mu}, A_{15 \mu}, B''_{ \mu}\right)$, the respective squared mass matrix is given by
{\footnotesize
\bea && M_{\mathrm{mass}}^{2NG}  =  \fr{g^2}{4} \crn
&&\left(%
\begin{array}{cccc}
u^2 + v^2 & \fr{1}{\sqrt{3}}(u^2 - v^2 ) &\fr{1}{\sqrt{6}}(u^2 - v^2 ) & \fr{t}{\sqrt{2}}(X_{\Phi_4 }u^2 -X_{\Phi_3 }v^2) \\
 & \fr 1 3 (u^2 + v^2 + 4\om^2) & \fr {1}{ 3 \sqrt{2}}  (u^2 + v^2 - 2 \om^2)  &\fr{t}{\sqrt{6}} (X_{\Phi_4 }u^2 +X_{\Phi_3 }v^2  - 2X_{\Phi_2 }\om^2)\\
& & \fr 1 6 (u^2 + v^2 + \om^2 + 9 V^2) &\fr{t}{2\sqrt{3}} (X_{\Phi_4 }u^2 +X_{\Phi_3 }v^2  + X_{\Phi_2 }\om^2 - 3X_{\Phi_1 }V^2)\\
& & &\fr{t^2}{2} (X^2_{\Phi_4 }u^2 +X^2_{\Phi_3 }v^2  + X^2_{\Phi_2 }\om^2  +X^2_{\Phi_1 }V^2)
\end{array}\,
\right).\crn
\label{m4eq37}
\eea
}
Following the above assumption, the SSB  following the pattern
\[ SU(4)_L\otimes U(1)_X   \stackrel{V}\longrightarrow
SU(3)_L\otimes U(1)_N   \stackrel{\om}\longrightarrow    SU(2)_L\otimes U(1)_Y
\stackrel{u, v}\longrightarrow   U(1)_Q\,
\]
will be used for constructing the matching relation of the gauge couplings and $U(1)$ charges of the group $SU(4)_L\times U(1)_X$
and those of the SM gauge group $SU(2)_L\times U(1)_Y$.  Corresponding to each  step of the breaking,
the neutral gauge boson states will be changed as follows:
\bea   && SU(4)_L\otimes U(1)_X   \stackrel{A_3,A_8,A_{15},B''}\longrightarrow
SU(3)_L\otimes U(1)_N   \crn
&& \stackrel{A_3,A_8,B',Z''_4}\longrightarrow    SU(2)_L\otimes U(1)_Y
\stackrel{A_3,B,Z'_3,Z'_4}\longrightarrow   U(1)_Q: A, Z,Z'_3,Z'_4\, . \eea
At the first  step of breaking, the  nonzero VEV $V\neq 0$ just results in $M_{\mathrm{mass}}^{2NG}\rightarrow M_{43}^{2}=M_{\mathrm{mass}}^{2NG}|_{w=v=u=0}$, where  the $M_{43}^{2} $ is
%--
\bea M_{43}^{2}=\fr{g^2}{4}\left(
                  \begin{array}{cccc}
                    0 & 0 & 0 & 0 \\
                    0 & 0 & 0 & 0 \\
                    0 & 0 & \frac{3 V^2}{2} & \frac{-3 ctV^2}{4\sqrt{2}} \\
                    0 & 0 & \frac{-3 ctV^2}{4\sqrt{2}} &\frac{3 c^2t^2V^2}{16} \\
                  \end{array}
                \right).
\label{M43} \eea
The transformation $C_{43}$  relating the two before- and after-breaking bases, namely
$(A_3,A_8,A_{15}, B'')^T= C^T_{43}\times(A_3,A_{8}, B',Z''_{4})^T$,  is given by
\bea C_{43}=\left(
              \begin{array}{cccc}
                1 & 0 & 0 & 0 \\
                0 & 1 & 0 & 0 \\
                0 & 0 & c_{43} & s_{43} \\
                0 & 0 & -s_{43} & c_{43} \\
              \end{array}
            \right)
 \label{C43}\eea
with
\be c_{43}\equiv \frac{ct}{\sqrt{8+c^2t^2}} \hs \mathrm{and}\; s_{43}=\frac{2\sqrt{2}}{\sqrt{8+c^2t^2}}. \label{43mix}\ee
After this step, only $A_{15}$ and $B''$ mix with each other to create the $U(1)_N$ gauge boson, denoted as $B'$,  of the $SU(3)_L\times U(1)_N$ group. Also,  the case  $c=0$ leads to $c_{43}$=0 and $s_{43}=1$, implying that the $SU(4)_L$ decouples from the $U(1)_X$. The diagonal squared mass matrix can be found as $M^2_{43d}=C_{43}M_{43}^{2}C^T_{43}=\mathrm{diagonal}\left(0,0,0,\frac{3 V^2}{2s^2_{43}}\right)$, including three massless and one massive values.

Similarly, the second  step of the breaking from $SU(3)_L\times U(1)_N\rightarrow SU(2)_L\times U(1)_Y$ can be done by the second transformation $C_{32}$ satisfying $(A_3,A_{8}, B',Z''_{4})^T=C^T_{32}\times (A_3,B,Z''_3,Z''_4)^T$, where $Z''_3$ and $Z''_4$, at this moment, are not mass eigenstates. The squared mass matrix now is $M_{\mathrm{mass}}^{2NG}\rightarrow M_{42}^{2}=M_{\mathrm{mass}}^{2NG}|_{v=u=0}$ so that $(A_3,A_{8}, A_{15},B'')^T=C^T_{42}\times (A_3,B,Z''_3,Z''_4)^T $. The concrete transformation of the two steps of the breaking are $C_{42}=C_{32}.C_{43}$. These transformations are given as follows,
%--
\bea
C_{32}=\left(
              \begin{array}{cccc}
                1 & 0 & 0 & 0 \\
                0 &  c_{32}&s_{32} & 0 \\
                0 &-s_{32} &c_{32} &0  \\
                0 & 0 & 0 & 1 \\
              \end{array}
            \right), \hs
%---
 C_{42}=\left(
              \begin{array}{cccc}
                1 & 0 & 0 & 0 \\
                0 &  c_{32}&c_{43}s_{32} & s_{43}s_{32} \\
                0 &-s_{32} &c_{43}c_{32} & s_{43}c_{32}  \\
                0 & 0 & -s_{43} & c_{43} \\
              \end{array}
            \right),
 \label{C432}\eea
%--
where
\be  s_{32} = \frac{2\sqrt{2}}{\sqrt{b^2t^2s^2_{43}+8}}=\frac{\sqrt{8+c^2t^2}}{\sqrt{8+(b^2+c^2)t^2}} , \hs c_{32}=\frac{bts_{43}}{\sqrt{b^2t^2s^2_{43}+8}}=\frac{bt}{\sqrt{8+(b^2+c^2)t^2}}.   \label{sc32}\ee
%--
After  two  steps of breaking, the squared mass matrix
\bea
M^{\prime 2}_{42}= C_{42}M^2_{42}C^T_{42}= \frac{g^2}{4}\left(
                  \begin{array}{cccc}
                    0 & 0 & 0 & 0 \\
                    0 & 0 & 0 & 0 \\
                    0 & 0 & \frac{4 w^2}{3s^2_{32}} & \frac{\sqrt{2}\left( -1+c_{43}s_{43}bt\right)w^2}{3s_{43}s_{32}} \\
                    0 & 0 &  \frac{\sqrt{2}\left( -1+c_{43}s_{43}bt\right)w^2}{3s_{43}s_{32}}  & \frac{\left( -1+c_{43}s_{43}bt\right)^2w^2 +9V^2}{6s^2_{43}}  \\
                  \end{array}
                \right) \label{m42p}
 \eea
 contains  two massless eigenvalues corresponding to the $A_3$ and $B$ gauge bosons of the SM gauge group $SU(2)_L\times U(1)_Y$. The two states $Z''_3$ and $Z''_4$ are still not eigenstates, but they decouple from the SM gauge bosons. They can be easily diagonalized, but that will be done later. To find the matching condition, we pay attention to the $B_{\mu}$ state involved with the neutral  part of the covariant derivative  that changes following the the steps of breaking, namely
 \bea D^{(41)}_{\mathrm{neutral}\mu}&=& \partial_{\mu}-i g\left( A_{3\mu}T^3+ A_{8\mu}T^8 + A_{15\mu}T^{15}+t\times X \times B''_{\mu} T^{16} \right)\crn
 \rightarrow  D^{(21)}_{\mathrm{neutral}\mu}&=& \partial_{\mu}-ig\left[A_{3\mu}T^3+ c_{32}B_{\mu}T^8 +c_{43} s_{32}B_{\mu}T^{15}+t\times X \times s_{43}s_{32}B_{\mu} T^{16} \right] \crn
 &=& \partial_{\mu}-igA_{3\mu}T^3 - i\frac{gt}{\sqrt{8+(b^2+c^2)t^2}}  B_{\mu}\left(bT^8+c T^{15}+XI_4 \right), \label{D41_21}\eea
 where all $ A_{8,15}$ and $B''$ are replaced by $B$ based on (\ref{C432}).  Identifying the $D^{(21)}_{\mu}$ in (\ref{D41_21}) with the covariant derivative defined in the SM, we derive that the gauge coupling of the $SU(4)_L$ is the $SU(2)_L$ coupling $g$. The other  two important equalities are
  \be   \frac{gt}{\sqrt{8+(b^2+c^2)t^2}} =g_1,  \hs  \mathrm{ and}\,  \;
    \frac{\widehat{Y}}{2} = bT^8+c T^{15}+XI_4,  \label{matchingSM}\ee
%--
where $g_1$ and $\widehat{Y}$ are the   coupling and charge operator of the SM $U(1)_Y$ gauge group.  The second formula in (\ref{matchingSM}) is consistent with the identification of  $\widehat{Y}$  from the definition of the electric charge operator (\ref{m4eq1}). Furthermore,  it can be seen that $\widehat{N}\equiv cT^{15}+X$ and $b\equiv\beta/\sqrt{3}$ are relations between  parameters defined in the gauge groups $SU(4)_L\times U(1)_X$ and $SU(3)_L\times U(1)_N$.

From $ g_1/g=s_W/c_W$, where $s^2_W=0.231$,  we find
\be t=\frac{g'}{g}=\frac{2\sqrt{2}s_W}{\sqrt{1-(1+b^2+c^2)s^2_W}}. \label{ft}\ee
 We emphasize  the ratio of two couplings - the parameter $t$ is  symmetric in changing from
 $q$ to $q'$. To see this, with the help of Eq. (\ref{m4eq3}),  the above formula can be written in terms of $q$ and $q'$ as follows:
\be t
 =  \frac{2\sqrt{2}s_W}{\sqrt{1-\left[ q + q'  - q q' + \fr 3 2 (1+q^2+q^{' 2})\right] s^2_W}}\, .
 \label{ftt}\ee
  Keeping in mind that  $s^2_W\simeq 0.23$, from (\ref{ftt}) we get a  constraint on electric charges of the new exotic leptons $E_a^q $  and $ E_a^{\prime q'}$, namely
\be q + q'  - q q' + \fr 3 2 ( 1+q^2+q^{' 2}) \leq 4 \, .
 \label{ftt1}\ee
The squared mass matrix (\ref{m42p}) can be diagonalized by a matrix $C'_{32}$, which gives  two mass eigenstates $Z'_3$ and $Z'_4$ in the second  step of the breaking. While the breaking from $SU(2)_L\times U(1)_Y$ to $U(1)_Q$ is taken by the well-known transformation $C_{21}$, the two transformative matrices are
\bea C_{21}= \left(
               \begin{array}{cccc}
                 s_W & c_W & 0 & 0 \\
                 c_W & -s_W & 0 & 0 \\
                 0 & 0 & 1 & 0 \\
                 0 & 0 & 0 & 1 \\
               \end{array}
             \right), \hs C'_{32}= \left(
               \begin{array}{cccc}
                1 & 0 & 0 & 0 \\
                0 & 1 & 0 & 0 \\
                 0 & 0 & c_{\alpha} & s_{\alpha} \\
                 0 & 0 & -s_{\alpha} & c_{\alpha} \\
               \end{array}
             \right),
  \label{c21}\eea
where $c_{\alpha}=\cos\alpha$ and $s_{\alpha}=\sin\alpha$  satisfy
\be t_{2\alpha}\equiv -\frac{2\sqrt{2}\left( -1+c_{43}s_{43}bt\right)w^2}{s_{43}s_{32}\left[\frac{\left( -1+c_{43}s_{43}bt\right)^2w^2 +9V^2}{2s^2_{43}} -\frac{4 w^2}{s^2_{32}}\right]}= \frac{4\sqrt{2}s_{43}s_{32}\left(-1+c_{43}s_{43}bt\right)w^2}{8s^2_{43}w^2-s^2_{32}\left[(-1+c_{43}s_{43}bt)^2w^2+9V^2\right]}.  \label{t2al}\ee
%--
Then we have  $M^2_{42d}=C'_{32}M^2_{42}C'^T_{32}=\mathrm{diag}(0,0,m^2_{Z'_3},m^2_{Z'_4})$, where
\bea m^2_{Z'_3}&=&\left(M^2_{42d}\right)_{33}= \frac{g^2}{4}\left[\frac{3s^2_{\alpha}V^2}{2s^2_{43}}+  \frac{\left[2\sqrt{2}c_{\alpha}s_{43}+s_{\alpha}s_{32}\left( -1+bs_{43}c_{43}t \right)\right]^2w^2}{6s^2_{43}s^2_{32}}\right],\crn
m^2_{Z'_4}&=&\left(M^2_{42d}\right)_{44}=  \frac{g^2}{4}\left[\frac{3c^2_{\alpha}V^2}{2s^2_{43}}+  \frac{\left[2\sqrt{2}s_{\alpha}s_{43}-c_{\alpha}s_{32}\left( -1+bs_{43}c_{43}t \right)\right]^2w^2}{6s^2_{43}s^2_{32}}\right].   \label{m2zp34}\eea
%--
The total transformation after all steps of breaking is $C=C_{21}.C'_{32}.C_{32}.C_{43}$.
The  squared mass matrix of the neutral gauge boson  transformed  under this rotation is derived as
\be M^2_{41}=C.M^{2NG}_{\mathrm{mass}}.C^T= \mathrm{diag}(0,0,m^2_{Z'_3},m^2_{Z'_4})+\delta{M^2_{41}},
 \label{m241} \ee
where $\delta{M^2_{41}}$ is a $4\times 4$ matrix having the property that  $\left(\delta{M^2_{41}}\right)_{ij}= \mathcal{O}(m_W^2)$ with all $i,j=1,2,3,4$. In addition, $\left(\delta{M^2_{41}}\right)_{i0}=\left(\delta{M^2_{41}}\right)_{0i}=0$ with any $i=1,2,3,4$, and $\left(\delta{M^2_{41}}\right)_{22}= m_Z^2$.   We can approximately  consider $M^2_{41}$ as the diagonal matrix, where $C$ is the transformation  relating the original and physical bases of neutral gauge bosons $(A_3,A_8,A_{15}, B'')^T$ and $(A,Z, Z_3,Z_4)^T$; precisely
%--
\bea A_{\mu} &=& s_W A_{3\mu}+c_W\left( c_{32}A_{8\mu}+c_{43}s_{32}A_{15\mu}+s_{43}s_{32}B''_{\mu} \right), \crn
Z_{\mu} &\simeq& c_W A_{3\mu}-s_W\left( c_{32}A_{8\mu}+c_{43}s_{32}A_{15\mu}+s_{43}s_{32}B''_{\mu} \right),\crn
 Z_{3\mu} &\simeq&Z'_{3\mu}= -s_{32}c_{\alpha}A_{8\mu}+\left(c_{43}c_{32}c_{\alpha}-s_{43}s_{\alpha}\right)A_{15\mu}+ \left(s_{43}c_{32}c_{\alpha}+c_{43}s_{\alpha}\right)B''_{\mu},\crn
 Z_{4\mu} &\simeq&Z'_{4\mu}= s_{32}s_{\alpha}A_{8\mu}-\left(c_{43}c_{32}s_{\alpha}+s_{43}c_{\alpha}\right)A_{15\mu}+ \left(c_{43}c_{\alpha}-s_{43}c_{32}s_{\alpha}\right)B''_{\mu}. \label{massev} \eea

Now we return to the Higgs content of the model. From the above presentation we explicitly  see that:
\ben
\item If $q, q' \neq 0\, ,\hs  q, q' \neq -1$, and $q\neq q'$, then the Higgs sector is smallest containing only four neutral Higgs fields.
\item  The case $q=q' =0$ has been considered in  \cite{noexotic},  where the Higgs sector contains ten neutral Higgs fields, and
 there are three non-Hermitian neutral gauge. This case is extremely complicated.
\item The case $q = q' =-1$  has been  considered in  \cite{anyqq1,anyqq,noexotic,cog,qqn1,qqn1k,ann}.
\item The case $q =  0,\,  q' = 1$  has been  considered in  \cite{flt,pp,qn1qo}.
\item The case $q= -1, q'= 0$  has been  considered in \cite{qn1qo}.
\item SU(4)(L) x U(1)(X) models with a little Higgs have been  presented in \cite{lit341}.
\een
Let us summarize the SSB pattern.  At the first step of symmetry breaking through $V$, only the following fields get masses:  the prime fermions including the exotic leptons $E’_i$ , quarks $T’$ and quarks $D’_\alpha$; and the gauge bosons $W_{34}$ and $Z’’$. At the second step of  SSB through $\om$, all remain  exotic  fermions - exotic leptons and quarks get masses. The charged gauge bosons in the  top right corner of the non-Hermitian gauge boson matrix [see,  (\ref{m4eq25})]  and the extra $Z^\prime$  obtain masses. Finally, the last step is possible through $u$ and $v$ and  all the SM fermions and  gauge bosons get masses.

%them dot 1---------
Let us take time to briefly review the development of the $\mbox{SU}(4) \otimes \mbox{U}(1)$ models. To our knowledge,
the first attempt by Fayyazuddin and Riazuddin \cite{fa1} introduced the decuplet. With the electric  charges of leptons  $q=0, \, q' =1 $, the limit on the sine-squared of the Weinberg angle was obtained: $\sin^2 \theta_W = 0.25$ and the bound for unification mass is as follows: $3.3 \times 10^4 \geq  m_X \geq  6.4 \times 10^3 $ GeV. At that time, the particle arrangement in Ref. \cite{fa1} was not correct.  The next step belongs to  M. B. Voloshin \cite{o341} who attempted to solve problem related to the realizability of the small mass and large magnetic moment of neutrinos. For this
purpose, the author focused on the lepton sector  only where the particle arrangement is the same as ours (see below).

The 3-4-1 model in the form discussed here was  proposed in \cite{flt,pp}.  The questions concerning anomaly cancellation
  and quantization of electric  charge, and the neutrino and generation nonuniversality were addressed in \cite{afree} and  \cite{pisanoc}, respectively.
In \cite{pisil}, the neutrinos and electromagnetic gauge invariance were discussed, while  the Majoron emitting neutrinoless double beta decay  in the minimal 3-4-1  model with right-handed neutrino containing a decuplet  were addressed in \cite{pisil,pisanoshel}. In association with discrete $Z_2$ symmetry, the model without exotic electric charges providing   a consistent mass spectrum, was proposed in \cite{ponce2004}.
 The $\mbox{SU}(4)_{(EW)} \times \mbox{U}(1)_{(B-l)}$  model with  left-right symmetry
  was proposed in \cite{341bl}.  It is interesting that the electroweak unification of quarks and leptons in a gauge group $\mbox{SU}(3)_C\times \mbox{SU}(4)\times \mbox{U}(1)$ was  built  in \cite{uquarklepton}. The muon anomalous magnetic moment in the $\mbox{SU}(4)\times \mbox{U}(1)_N$ model was  considered in \cite{mhta}. The neutrino mass and mixing in the special formalism were presented in  \cite{palcu}.

It is worth mentioning that, except for the  suspersymmetric 3-4-1 model \cite{susy341},
the Higgs potential containing a decuplet is  presented for the first time  in this paper.
%---------het

Now we turn to the  model similar to the one originally built in  \cite{flt,pp} with $q=1, q'=0$.
 \section{\label{m341rhn}Minimal 3-4-1 with right-handed neutrinos}
 %-------
Let us consider a model in which leptons are  arranged as
\be f_{a L} = ( \nu_a\, , l_a \, , l_a^{c} \, , \nu^c_a )_L^T \sim \left(1,4, 0\right) ,\hs  a = e, \mu, \tau \, ,
\label{m4eq388}
\ee
where we have in mind that $\nu^c_{ L} \equiv  (\nu_{ R})^c $  and the charge conjugation of $ f_{a L}$:
$f^c_{aR}\equiv \left( f_{aL}\right)^c=(\nu^c_{aR}\;, l^c_{aR},\; l_{aR},\; \nu_{aR})^T$.

%------
One  quark generation is  in quadruplet:
\bea Q_{3 L} &  = & \left(%
%\begin{array}{c}
u_3  \, ,
d_3 \, ,
T \, ,
T^\prime
%\end{array}\,
\right)^T_L \sim \left(3, 4,   \fr{2}{3} \right)\, , \crn
u_{3 R} & \sim & (3, 1, 2/3 )\, , \hs  d_{3 R}  \sim  (3, 1, - 1/3 )\, , \crn
T_R & \sim &  \left(3, 1, \fr{5}{3}\right) \, , \hs T^\prime _R \sim \left(3, 1, \fr{2}{3}\right).
\label{m4eq458}
\eea
The exotic quarks have electric charges: $q_T = \fr 5 3$, $q_{T^\prime} =  \fr{2}{ 3}$.
Two other quark generations are  in antiquadruplet
%-----
\bea Q_{\al L} &  = & \left(%
%\begin{array}{c}
d_\al  \, ,
- u_\al \, ,
D_\al \, ,
D_\al^\prime
%\end{array}\,
\right)^T_L \sim \left(3, 4^*,  - \fr{1}{3} \right)\, , \hs \al = 1, 2 , \crn
u_{\al R} & \sim & (3, 1, 2/3 )\, , \hs  d_{\al R}  \sim  (3, 1, - 1/3 )\, , \crn
D_{\al R} & \sim &  \left(3, 1, -\fr{4}{3}\right) \, , D_{\al R}^\prime  \sim \left(3, 1,- \fr{1}{3}\right).
\label{m4eq488}
\eea
The exotic quarks have electric charges:
 $q_{D_\al} = - \fr 4 3$, $q_{D^\prime_{\al}} =  - \fr{1}{3}$.

Applying Eq.(\ref{m4eq1}) to Eq.(\ref{m4eq388}),
we obtain
\be b = -\sqrt{ 3 }\, , \hs c = 0\, , \hs X_{f_{a L} } = 0.
\label{m4eq395}
\ee
Then the electric charge operator, for the quadruplet, has the form
\be
Q
= \textrm{diag}\left(%
 X \,  ,  -1 + X \, , 1 +X \, ,X
\right)\, . \label{m4eq408}
\ee
For SSB,  we need four Higgs quadruplets, namely,
\bea \chi  & =  &  \left(%
%\begin{array}{c}
\chi_1^{0} \, ,
\chi_2^{-}\, ,
\chi_3^{+}\, ,
\chi_4^0
%\end{array}\,
\right)^T \sim \left(1, 4, 0\right)\, , \hs \ph =  \left(%
%\begin{array}{c}
\ph_1^{-} \, ,
\ph^{--}\, ,
\ph^0\, ,
\ph_2^{-}
%\end{array}\,
\right)^T \sim \left(1, 4,  - 1\right)\, ,\crn
 \rho  & =  &  \left(%
%\begin{array}{c}
\rho_1^{+} \, ,
\rho^0\, ,
\rho^{++}\, ,
\rho_2^{+}
%\end{array}\,
\right)^T \sim \left(1, 4, 1\right)\, , \hs \eta =  \left(%
%\begin{array}{c}
\eta_1^{0} \, ,
\eta_2^{-}\, ,
\eta_3^{+}\, ,
\eta_4^{0}
%\end{array}\,
\right)^T \sim \left(1, 4,  0 \right)\, .
\label{m4eq425}
\eea
In \cite{ann}, the  Higgs sector contains only three Higgs quadruplets, but to produce masses of charged leptons and neutrinos,
the nonrenormalizable effective dimension-five and -nine operators were used. Here we prefer the  original  way   in \cite{flt,pp}.

The Yukawa couplings for the quark sector are
\bea- L^q_{\mathrm{Yukawa}}  & = & h^{t}\, \overline{Q_{3 L} }\,  \eta u_{3 R}   + h^{b}\, \overline{Q_{3 L} } \rho d_{3 R}   +
h^{T}\overline{Q_{3 L} }\,  \ph \,  T_{R}   + h^{T^\prime}\, \overline{Q_{3 L} }\,  \chi\,  T^\prime_{ R}    \crn
& + &  h^{d2}_{\al \bet}\overline{Q_{\al  L} } \eta^\dag d_{\bet R}   +
h^{u2}_{\al \bet}\overline{Q_{\al  L} } \rho^\dag u_{\bet R}
+  h^{D2}_{\al \bet}\overline{Q_{\al  L} } \ph^\dag D_{\bet R}   +
h^{D'2}_{\al \bet}\overline{Q_{\al  L} } \chi^\dag D^\prime_{\bet R}
\crn&+&  \mathrm{H. c.}.
\label{m4eq468}
\eea
If  the Higgs sector has VEV structure as
\bea \langle \chi \rangle & =  &  \left(%
%\begin{array}{c}
0 \, ,
0\, ,
0\, ,
\fr{V}{\sqrt{2}}
%\end{array}\,
\right)^T \, , \hs
 \langle \ph \rangle  =    \left(%
%\begin{array}{c}
0 \, ,
0\, ,
\fr{\om}{\sqrt{2}}\, ,
0
%\end{array}\,
\right)^T ,\crn
\langle \rho \rangle & =  &  \left(%
%\begin{array}{c}
0\, ,\fr{v}{\sqrt{2}}\, ,
0\, , 0
%\end{array}\,
\right)^T \, , \hs
\langle \eta \rangle  =    \left(%
%\begin{array}{c}
\fr{u}{\sqrt{2}}\, ,
0\, , 0\, ,
0
%\end{array}\,
\right)^T,
\label{l63}
\eea
then the quarks get masses as follows:
\bea m_{u_3} & = & h^t \fr{u}{\sqrt{2}}, \,  m_{d_3} = h^b \fr{v}{\sqrt{2}} \, ,
m_{T} = h^T \fr{\om}{\sqrt{2}}\, , m_{T^\prime} = h^{T^\prime} \fr{V}{\sqrt{2}}\, ,\crn
(m_{d_2})_{ \al \bet  } & = & h^{d2}_{\al \bet} \fr{u}{\sqrt{2}}\, ,  (m_{u_2})_{ \al \bet } = -h^{u2}_{\al \bet} \fr{v}{\sqrt{2}}\, ,
 (m_{D_2})_{ \al \bet} = h^{D2}_{\al \bet} \fr{\om}{\sqrt{2}}\, ,  (m_{D^ \prime_2})_{ \al \bet } =
 h^{D'2}_{\al \bet}\ \fr{V}{\sqrt{2}}\, .
\label{m4eq506}
\eea
Until now, the leptons were massless.
To produce masses for leptons from the renormalizable Yukawa interactions, we will base it on the product
$\overline{f_{a L} }  f^c_{b R} \sim {6_A \oplus 10^*_S}$.  If an  antisymmetric ${\bf 6}$ is used,  then the lepton mass
 matrix will be antisymmetric; consequently one lepton is still massless. So, the  better way is to
  introduce a symmetric decuplet (${\bf 10}_S$) given by
\be H' \sim (1, {\bf 10}, 0)  =\left(%
\begin{array}{cccc}
H_1^0 &H_1^-  & H_2^+& H_2^0 \\
 H_1^- & H_1^{--}     & H_3^0 &  H_3^- \\
H_2^+ & H_3^0  & H_2^{++}&H_4^+ \\
H_2^0 &H_3^-&H_4^+&H_4^0\
\end{array}\,
\right)\, . \label{Higg10}
\ee
The gauge-invariant  Lagrangian of the $10$-plet is given by
\be L_0^{H'} = \mathrm{Tr }\left[ (D_\mu H')^\dag  D^\mu H'\right] -  V.
 \label{Higg11}
\ee
We will show that  the Higgs content written in (\ref{Higg10}) should be redefined.
To clarify this, let us consider the kinetic part of $H'$ in (\ref{Higg11})
\bea L^{H'}_{\mathrm{kinetic}} & = & \textrm{Tr}\, \left[ (\pa_\mu H')^\dag  \pa^\mu H'\right] \crn
& = & \left[\pa_\mu H_1^{0*} \pa^\mu H_1^0 + \pa_\mu H_4^{0*} \pa^\mu H_4^0
+ \pa_\mu H_1^{++}\pa^\mu  H_1^{--}
 + \pa_\mu H_2^{++} \pa^\mu H_2^{--}\right.  \crn
& & + 2( \pa_\mu H_2^{0*}\pa^\mu  H_2^0 + \pa_\mu H_3^{0*}\pa^\mu  H_3^0 \crn
& &
+\left.  \pa_\mu H_1^+\pa^\mu  H_1^- + \pa_\mu H_2^+ \pa^\mu H_2^-
+ \pa_\mu H_3^+ \pa^\mu H_3^- + \pa_\mu H_4^+ \pa^\mu H_4^- )\right].
\label{Higg61}
\eea
The factor 2 in the second line of  (\ref{Higg61}) shows   that nondiagonal fields in (\ref{Higg11})
 must be redefined as follows:
\be H' \rightarrow H  = \fr{1}{\sqrt{2}}\left(%
\begin{array}{cccc}
\sqrt{2} H_1^0 &H_1^-  & H_2^+& H_2^0 \\
 H_1^- & \sqrt{2} H_1^{--}     & H_3^0 &  H_3^- \\
H_2^+ & H_3^0  &\sqrt{2} H_2^{++}&H_4^+ \\
H_2^0 &H_3^-&H_4^+&\sqrt{2} H_4^0\
\end{array}\,
\right)\, . \label{Higg62}
\ee
If so, we have
\bea  \textrm{Tr} \,\left[ ( H)^\dag   H\right]
& = & H_1^{0*}  H_1^0 +  H_4^{0*}  H_4^0  +  H_2^{0*}  H_2^0 +  H_3^{0*}  H_3^0\crn
& &+  H_1^{++}  H_1^{--}
 +  H_2^{++}  H_2^{--}
+ H_1^+  H_1^- +  H_2^+  H_2^-
+  H_3^+  H_3^- +  H_4^+  H_4^-.
\label{Higg63}
\eea
  In what  follows,  we will use $H$ only.

The Yukawa interaction for the lepton  is given by
\bea - L_{\mathrm{Yukawa}}^{l} & = & h^l_{a b}  \overline{f_{a L} } H  f^c_{b R}  +\mathrm{H. c.}\crn
%---
&=& \fr{h^l_{a b}}{\sqrt{2}} \left[   \overline{\nu_a}_L \left( \sqrt{2}\nu^c_{b R} H_1^0 + l^c_{b R} H_1^- + l_{bR} H_2^+
+ \nu_{b R} H_2^0\right)
\right. \crn
&+& \overline{l_{aL}} \left(\nu^c_{bR} H_1^- +\sqrt{2} l^c_{bR} H_1^{--}  + l_{b R} H_3^0 + \nu_{b R} H_3^- \right)
\crn
&+& \overline{l^c_{aL}} \left(\nu^c_{b R} H_2^+ + l^c_{b R} H_3^{0}  + \sqrt{2}l_{bR} H_2^{++} + \nu_{b R} H_4^+ \right)
\crn
&+& \left. \overline{\nu^c_{aL}} \left(\nu^c_{b R} H_2^0 + l^c_{b R} H_3^{-}  + l_{b R} H_4^{+} +\sqrt{2}
\nu_{bR} H_4^0 \right)\right]+\mathrm{ H. c.}.
\label{m4eq4358}
\eea
 As usual,  assuming an expansion of the neutral Higgs fields as follows,
 \be H^0_3 = \fr{v' + R_{H_3^0} - i I_{H_3^0} }{\sqrt{2}}\, , \hs H^0_2 = \fr{\ep + R_{H_2^0} - i I_{H_2^0} }{\sqrt{2}}\, ,
 \label{lm81}
\ee
then
 \be  \langle H \rangle  =\fr{1}{2}\left(%
\begin{array}{cccc}
0 & 0 &  0 & \epsilon \\
 0 & 0   & v' &  0 \\
0 & v'  & 0 &0 \\
\epsilon &0 &0&0\
\end{array}\,
\right)\, . \label{Higg14}
\ee
%--------
The charged leptons get mass matrix  given by
\bea (m_{l})_ {a b} & = & \fr{h^{l}_{a b}}{\sqrt{2}} \langle H_3^0 \rangle = \fr{h^{l}_{a b} \, v'}{2}
   \, .\label{m4eq448}
\eea
The neutrinos obtain  the  Dirac mass  by $\langle H_2^0 \rangle$  in the same Yukawa coupling  matrix:
\bea (m_{\nu})_{ a b} & = &  \fr{h^{l}_{a b}}{\sqrt{2}} \langle H_2^0 \rangle = \fr{h^{l}_{a b}\,
 \epsilon}{2}.
   \,\label{m4eq448a}
\eea
The neutrino Majorana mass will follow from
$\langle H_1^0 \rangle$ and $\langle H_4^0 \rangle$.

Noting that the mixing parameters among the charged leptons are tiny, while those among the neutrinos are large,
the matrix in (\ref{m4eq448a}) must be modified. It is hoped that the radiative corrections
will provide the mixing that matches  the  current experimental data \cite{data}.  We will return to this problem
in our future work.

It is emphasized that there are flavor-lepton violating interactions in Eq. (\ref{m4eq4358}).

The lepton number operator is constructed from the diagonal  generators as follows,
\be L = \al T_3 + \bet T_8 + \ga T_{15} + {\cal L}.
\label{m4eq60}
\ee
For the general case, let us assume that the new extra leptons $E$ and $E'$ acquire lepton number $l$ and $l'$, respectively,
\be f_{a L} = ( \nu_a\, , l_a \, , E_a \, , E_a' )_L^T  ,\hs  a = e, \mu, \tau \, .
\label{m341may131}
\ee
Applying (\ref{m4eq60}) for (\ref{m341may131}), we obtain
\be \al = 0\, ,\; {\cal L}_{f_{a L}} = \frac{1}{2} + \fr  1 4 (l + l')\, ,
\bet =\fr {2(1-l)}{ \sqrt{3}}\, , \ga =\fr{(2 + l - 3 l')}{ \sqrt{ 6}}.
\label{m341may132}
\ee
The vanishing of the coefficient  $\al$   is a consequence of the lepton number conservation in the SM. Thus,
\be L =  \fr {2(1-l)}{ \sqrt{3}}  T_8 + \fr{(2 + l - 3 l')}{ \sqrt{ 6}}  T_{15} + {\cal L}.
\label{m341may133}
\ee
The above formula is useful for the extensions where the flavor  discrete symmetries such as $A_4, S_3$,  etc, are implemented.

Now we return to our model,  where the  lepton quadruplet in (\ref{m4eq388})  contains $l_a^c$ and $\nu_a^c$
with lepton number $(-1)$. We  then get
\be \bet = \fr{4}{\sqrt{3}}\, , \ga =  \fr{2\sqrt{6}}{3}\, .
\label{m4eq61}
\ee
Hence, the lepton number operator in the minimal 3-4-1 model with right-handed neutrinos gets the form
\be L =  \fr{4}{\sqrt{3}} \left( T_8 + \fr{1}{\sqrt{2}} T_{15}\right) + {\cal L}\, .
\label{lm3}
\ee
This formula is an  extension of that in the 3-3-1 model \cite{lchang}.  For the quadruplet, this operator  has  the form
\be
 L = \textrm{diag}
 \left(%
1 + {\cal L}\, ,  1 + {\cal L} \, , - 1 + {\cal L} \, , -1 + {\cal L}
%\end{array}\,
\right).
\label{m4eq62}
\ee
The fields with  nonzero lepton numbers  are listed in Tables \ref{bcharge}, \ref{lnumberH}, and    \ref{lnumberF}.
\begin{table*}
\caption{
     ${\cal B}$ and ${\cal L}$ charges for the multiplets in
the 3-4-1 model with right-handed neutrinos.}
\begin{center}
\begin{tabular}{l|ccccc|cccccccc|c}
Multiplet & $\chi$ &$\ph$ & $\eta$ & $\rho$ & $H$ & $Q_{3L}$ & $Q_{\al L}$ &
$u_{aR}$&$d_{aR}$ &$T_R$ &$T^\prime_R$ & $D_{\al R}$ &$D^\prime_{\al R}$ & $f_{aL}$  \\
\hline $\cal B$ charge &$0$ &$0$ & $ 0  $ &
 $ 0  $ & $ 0  $ & $\fr 1 3  $ & $\fr 1 3  $& $\fr 1 3  $ &$\fr 1 3  $ &
 $\fr 1 3  $ &  $\fr 1 3  $& $\fr 1 3  $& $\fr 1 3  $&
 $0  $\\
\hline $\cal L$ charge &$1$ & $1$ &$-1 $ &  $- 1 $ &$ 0$ &
   $- 1  $ & $ 1 $&$ 0$ & $0$ & $-2$& $-2$& $2$&$2$&
 $ 0  $\\
\end{tabular}
\label{bcharge}
\end{center}
\end{table*}

\begin{table*}
\caption{
    Nonzero lepton number $L$ of the Higgs fields  in the 3-4-1 model with right-handed neutrinos.}
\begin{center}
\begin{tabular}{l|cc|cc|cc|cc|ccccccc}
    \hline
        Fields
&$\chi_1^0$&$\chi_2^-$&$\ph^-_1$
& $\ph^{--}$&$\rho^{++}$&$\rho_2^+$&$ \eta_3^+$&$ \eta_4^0 $ &$ H_1^0$&$ H_4^0 $&$ H_1^+$&$ H_4^+ $&$H_1^{++}$&$H_2^{++}$ \\
    \hline
        $L$ & $2$ & $2$ & $2$ & $2$
        &$-2$&$-2$&$-2$&$-2$&$2$&$-2$&$-2$&$-2$&$ -2$&$ -2$\\
    \hline
\end{tabular}
\label{lnumberH}
\end{center}
\end{table*}

\begin{table*}
\caption{
    Nonzero lepton number $L$ of fermion  in the 3-4-1 model with right-handed neutrinos.}
\begin{center}
\begin{tabular}{l|cc|ccccc}
    \hline
        Fields &  \, $l_a$ \;& $\nu_a$ \,  &
$T $&$T^\prime$&$ D_{\al}$ &$ D^\prime_{\al}$ \\
    \hline
        $L$ \, & \, $1$& $1$&$-2$ & $-2$ & $2$ & $2$
       \\
    \hline
\end{tabular}
\label{lnumberF}
\end{center}
\end{table*}

Now we turn to the gauge boson sector. The
contribution to the gauge boson masses from $H$ arises from a piece,
 \bea
{\cal{L}}^{H}_{\mathrm{mass}} & = &\mathrm{Tr}[(D_\mu\langle H \rangle)^+
(D^{\mu}\langle H \rangle)] \crn
& = & g^2 \mathrm{Tr} [ (P_\mu^{CC} \langle H \rangle)^\dag  (P^{\mu CC} \langle H \rangle) +
 (P_\mu^{NC} \langle H \rangle)^\dag  (P^{\mu NC} \langle H \rangle)],
  \label{Higgs12}\eea
 where
 \bea
D_\mu & = & \pa_\mu - i g\sum_{a=1}^{15} A_{a \mu} T_a  - i g' X B''_\mu T_{16} \, \crn
& \equiv & \pa_\mu - i g P_\mu \crn
& \equiv & \pa_\mu - i g P_\mu^{NC} - i g P_\mu^{CC}.
 \label{Higgs12t}
\eea
As a result of the symmetric form of the two quadruplets, we have (for details, see \cite{donglong})
\bea
(P_\mu  H) _{ij} & = &  (P_{\mu })_i^k  H _{k j} + (P_{\mu })_j^k  H _{k i}
\, . \label{Higg13}
\eea
For the gauge boson masses, one needs to calculate
\bea
(P_{\mu  } \langle H \rangle)_{ij} & = & (P_{\mu })_i^k\langle H \rangle_{k j} + (P_{\mu })_j^k \langle H \rangle_{k i}.
 \label{Higg13t}
\eea
We first deal with the charged gauge boson masses being defined through
\bea P_\mu^{CC} & = &\fr 1 2 \sum_{a} \la_a A_{a} \, , \hs a = 1,2,4,5,6,7,9,10,11,12,13,14\crn
&=&\fr{1}{ \sqrt{2}}\left(%
\begin{array}{cccc}
0 & W^{'+} &W_{13}^{-} & W_{14}^{0}\\
 W^{'-}  & 0 & W_{23}^{--}  &W_{24}^{-}\\
W_{13}^{+} & W_{23}^{++} & 0 &W_{34}^{+} \\
\left(W_{14}^{0}\right)^*&W_{24}^{+}&W_{34}^{-}&0
\end{array}\,
\right)_\mu
=\fr{1}{ \sqrt{2}}\left(%
\begin{array}{cccc}
0 & W^{'+} &Y^{'-}& N^{0}\\
 W^{'-}  & 0 & U^{--} &X^{'-}  \\
Y^{'+} & U^{++} & 0 &K^{'+}\\
\left(N^{0}\right)^*&X^{'+}& K^{'-}&0
\end{array}\,
\right)_\mu,
\label{m4eq258}
\eea
where we have denoted  $\sqrt{2}W_\mu^{'+ } \equiv A_{1\mu} - i A_{2\mu}$,
$Y^{'-} \equiv W_{13}^{-}\, , X^{'-} \equiv  W_{24}^{-}\, , K^{'+} \equiv W_{34}^{+}\, ,
 U^{--} \equiv W_{23}^{--} $ and $N^0 \equiv W_{14}^0$.

The masses of the non-Hermitian neutral $N^0$ and doubly charged $U^{\pm \pm}$ gauge bosons are
\be  m^2_{U^{\pm \pm}}  =  \fr{g^2( \om^2  + v^2 + 4  v'^2)}{4}\, , \hs
m^2_{N^0}   =   \fr{g^2( V^2 + u^2 +  4 \epsilon^2)}{4},\,
\label{may55a}
\ee
where the $N^0$ and $N^{0*}$ gauge bosons do not mix with other Hermitian neutral gauge bosons.
The squared mass matrix of the singly charged gauge bosons is rewritten  in the basis $(W^{'\pm},K^{'\pm},X^{'\pm},Y^{'\pm})^T$ as follows:
\bea M^2_{G^{\pm}}=\frac{g^2}{4} \left(
                                   \begin{array}{cccc}
                                     v^2+u^2+v''^2 & 2v'\epsilon & 0 & 0  \\
                                      & w^2+V^2+v''^2    &
                                    0 & 0 \\
                                      &  & v^2+V^2+v''^2 & 2v'\epsilon \\
                                      &  &  & u^2+w^2+v''^2 \\
                                   \end{array}
                                 \right),
 \label{m2singylGB}\eea
 where $v''^2\equiv v'^2+\epsilon^2$. In the limit $\epsilon=0$, all of  the above masses are the same as those given in \cite{pp}
 (with unique differences associated with $v''$,  since in \cite{pp}  the authors did not use the redefined decuplet).
 Note that all nondiagonal elements of the  matrix (\ref{m2singylGB}) are proportional to $v'\epsilon$, which are much  smaller than the diagonal ones;  therefore the  mass eigenstates of the singly charged gauge bosons can be
  reasonably identified with those given in   \cite{pp}.

 The physical states are determined as
\be W_\mu  =  \cos \theta \, W'_\mu - \sin \theta\, K'_\mu\, , \hs
K_\mu  =   \sin \theta\,  W'_\mu + \cos \theta\, K'_\mu\, ,
\label{l66}
 \ee
 where the $W-K$ mixing angle $\theta$ characterizing lepton number violation is given by
\be
\tan 2\theta  =   \fr{4v' \ep}{V^2 + \om^2 - u^2-v^2}\, .
\label{eq6}
\ee
For the $X-Y$ mixing, we obtain the physical states,
\be Y_\mu  =  \cos \theta' \, Y'_\mu - \sin \theta'\, X'_\mu\, ,\hs
X_\mu   =   \sin \theta'\,  Y'_\mu + \cos \theta'\, X'_\mu\,
\label{l67}
 \ee
with  the mixing angle $\theta'$  defined as
\be
\tan 2\theta'  =   \fr{4v' \ep}{V^2 - \om^2 - u^2 +v^2}\, .
\label{eq7}
\ee
The masses of the physical states are determined as
\bea m^2_{W^\pm} &\simeq & \fr{g^2}{4} (v^2+u^2+v''^2) \,  ,\hs m^2_{K^\pm}
 \simeq  \fr{g^2}{4} ( V^2+ w^2+v''^2) \,  ,\crn
m^2_{X^\pm} &\simeq & \fr{g^2}{4}(V^2+v^2+v''^2) \,  ,\hs m^2_{Y^\pm}
 \simeq  \fr{g^2}{4} (w^2+u^2+v''^2) \,  .
\label{lm82}
\eea
From (\ref{lm82}), it follows that
\be  v^2+u^2+v''^2 \simeq v^2_{\mathrm{SM}} = (246 \, \textrm{GeV})^2\, ,
\label{lm83}
\ee
and bounds on the singly charged gauge boson  mass splitting are
\be  | m^2_{K} - m^2_{X}| \leq m^2_{Y}\, , \hs | m^2_{K} - m^2_{X} -
 m^2_{Y}| \leq m^2_{W}\, .
\label{lm84}
\ee
Comparing (\ref{eq6}) with (\ref{eq7}), we see that the mixing angle between the lightest $W$ and heaviest $K$ is
smaller than the  $X-Y$ mixing angle.
 The mixing angle is quite small and can be  constrained from the $W$ decay width
  (as in the economical 3-3-1 model  \cite{ecn331c, eco}).

From the experimental point of view, the approximation in previous works \cite{pp,cog}, $V= \om$ creates
the difficulty in  distinguishing between  the bileptons  $X$ and  $Y$. So the natural way is to assume $V \gg \om$.
Note that $U^{\pm \pm}$ and $Y^\pm$ are   similar to the  singly charged gauge bosons  in the minimal 3-3-1 model \cite{331ppf}, while $N^0$ and  $X^\pm$  play the similar role  in the 3-3-1 model with right-handed neutrinos \cite{flt,331flt}. The heaviest singly charged
gauge bosons  $K^\pm$  are the  completely new ones that  couple with the exotic quarks and right-handed  leptons only
(see, Sec. \ref{current}). In our assignment (and also in Voloshin's paper \cite{o341}), particles  belonging to the minimal version are lighter than those in the 3-3-1 model with right-handed neutrinos
[i,e., Eqs.(\ref{may55a}] and (\ref{lm82})). For the original 3-4-1 model \cite{flt,pp}, the above consequence is the opposite.

%------Long sua cau duoi------------

Now we turn to the neutral gauge boson sector. In the basis $\left(A_{3 \mu}, A_{8 \mu}, A_{15 \mu}, B''_{ \mu}\right)$, the squared mass matrix for the neutral gauge bosons  is given by
{\footnotesize
\bea && M_{\mathrm{mass}}^{2NG}  =  \fr{g^2}{4}\crn
&\times&
\left(%
\begin{array}{cccc}
u^2 + v^2 + v''^2 & \fr{1}{\sqrt{3}}(u^2 - v^2+ v''^2) &\fr{1}{\sqrt{6}}(u^2 - v^2 - 2v''^2 ) &- \fr{t}{\sqrt{2}}v^2 \\
 & \fr 1 3 (u^2 + v^2 + 4\om^2+ v''^2) & \fr {1}{ 3 \sqrt{2}}  (u^2 + v^2 - 2 \om^2 - 2v''^2)
   &\fr{t}{\sqrt{6}} ( v^2  + 2\om^2)\\
& & \fr 1 6 (u^2 + v^2 + \om^2 + 9 V^2 + 4v''^2) &\fr{t}{2\sqrt{3}} ( v^2 - \om^2 )\\
& & &\fr{t^2}{2} ( v^2  + \om^2)
\end{array}\,
\right).
\label{may55}
\eea
}
 It has a property that $\mathrm{Det}(M_{\mathrm{mass}}^{2NG})=0$, implying a massless state of the photon.

 The mass eigenvalues of this matrix can be done the same way as in  the general case. For this particular
  case, we have
 \bea s_{43}&=&1,\;  c_{43}=0, \; s_{32}= \frac{2\sqrt{2}}{\sqrt{8+3t^2}},\;  c_{32}= \frac{-\sqrt{3}t}{\sqrt{8+3t^2}},\crn
  t_{2\alpha}&=& \frac{2\sqrt{8+3t^2}w^2}{9V^2-(7+3t^2)w^2}.\label{cTrans}\eea
 %--
 The transformative matrix $C_{41}$ satisfying $ (A,Z,Z'_3,Z'_4)^T=C_{41}.(A_3,A_8,A_{15}, B'')^T$ now has the form
 \bea C_{41}=\left(
               \begin{array}{cccc}
                 s_W & c_Wc_{32} & 0 & c_Ws_{32} \\
                 c_W & -s_Wc_{32} & 0 & -s_Ws_{32} \\
                 0 & -c_{\alpha}s_{32} & -s_{\alpha} & c_{\alpha}c_{32} \\
                 0 &s_{\alpha}s_{32} & -c_{\alpha} &- s_{\alpha}c_{32}   \\
               \end{array}
             \right).
  \label{TransRHN}\eea
  %---
 The squared mass matrix in the new basis is obtained as
 \bea M^{2NG}_{41}= \left(
                   \begin{array}{cccc}
                     0 & 0 & 0 & 0 \\
                      & m^2_{Z} &m^2_{23}  & m^2_{24} \\
                    &  &  m^2_{Z'_3} & m^2_{34} \\
                      &  & & m^2_{Z'_4} \\
                   \end{array}
                 \right),
  \label{m2NGRHNp}\eea
  where
    {\footnotesize
  \bea
  m^2_{Z}&=&\frac{g^2(v^2+u^2+v''^2)}{4c^2_W}=\frac{m_W^2}{c^2_W},\crn
  m^2_{Z'_3}&=& \frac{g^2}{24}\left[ 9s^2_{\alpha} V^2+\left(s_{\alpha}-c_{\alpha}\sqrt{8+3t^2}\right)^2w^2\right]+\frac{g^2}{24}\left[\left(\sqrt{2} c_{\alpha}s_{32} +s_{\alpha}\right)^2u^2\right.\crn
  &+&\left. \left(s_{\alpha} +\frac{(3t^2+4)c_{\alpha}s_{32}}{2\sqrt{2}}\right)^2v^2 +2\left(\sqrt{2} s_{\alpha} -c_{\alpha} s_{32}\right)^2v''^2  \right],\crn
   m^2_{Z'_4}&=&\frac{g^2}{24}\left[ 9c^2_{\alpha} V^2+\left(c_{\alpha}+s_{\alpha}\sqrt{8+3t^2}\right)^2w^2\right]+\frac{g^2}{24}\left[\left(c_{\alpha}-\sqrt{2} s_{\alpha}s_{32} \right)^2u^2\right.\crn
  &+&\left. \left(c_{\alpha} -\frac{(3t^2+4)s_{\alpha}s_{32}}{2\sqrt{2}}\right)^2v^2  +2\left(\sqrt{2} c_{\alpha} +s_{\alpha} s_{32}\right)^2v''^2 \right],\crn
  %--
  m^2_{23}&=&\frac{g^2}{4}\left[ -\left( \frac{c_{\alpha}\sqrt{1-4s^2_W}}{c^2_W\sqrt{3}} +\frac{s_{\alpha}}{\sqrt{6}c_W}\right)u^2 +\left( \frac{c_{\alpha}(3-2c^2_W)}{c^2_W\sqrt{3(1-4s^2_W)}}+\frac{s_{\alpha}}{\sqrt{6}c_W}\right)v^2\right.\crn
  &+&\left.\left( -\frac{c_{\alpha}\sqrt{1-4s^2_W}}{c^2_W\sqrt{3}}+\frac{\sqrt{2} s_{\alpha}}{\sqrt{3}c_W}\right)v''^2\right]\sim\mathcal{O}(m_W^2),\crn
  %---
   m^2_{24}&=&\frac{g^2}{4}\left[\frac{c_{\alpha}(v^2-u^2+2v''^2)}{\sqrt{6}c_W}+\frac{s_{\alpha}\left[(4c_W^2-3)(u^2+v''^2)+(2c^2_W-3)v^2 \right]}{c^2_{W}\sqrt{3(4c_W^2-3)}}\right]\sim\mathcal{O}(m_W^2),\crn
   %--
   m^2_{34}&=&\frac{g^2}{24}\left[ \frac{c_{2\alpha}}{\sqrt{3t^2+8}}\left[ 4(u^2-2v''^2)+(3t^2+4)v^2\right]\right.\crn
   &+ &\left. \frac{s_{2\alpha}}{2(3t^2+8)}\left[(3t^2-8)u^2-(8+21t^2+9t^4)v^2+4(3t^2+4)v''^2\right]\right]\sim\mathcal{O}(m_W^2).
  \label{eNGmass} \eea
  }
It can be seen that all of the nondiagonal elements are in the order of $\mathcal{O}(m_W^2)$. Therefore, they are much smaller than $m^2_{Z'_3}$ and $m^2_{Z'_4}$, implying that   these two values can be approximately the eigeinvalues of the matrix (\ref{m2NGRHNp}). Furthermore, the largest contribution to the $m^2_{Z}=\left(M^{2NG}_{41}\right)_{22}$ in the final diagonal matrix, which is proportional to $m^2_{W}\times \mathcal{O}(\frac{u^2+v^2}{V^2+w^2})$, is also small.  In conclusion, the matrix (\ref{m2NGRHNp}) can be considered as the diagonal matrix where the eigenvalues correspond to the diagonal elements, and the matrix $C_{41}$ is the one relating the two original and mass bases of the neutral gauge bosons.
%---
 %------
\subsection{\label{current}Currents}
From the Lagrangian of the fermion
\be L_{\mathrm{fermion}}  = i \sum_f \overline{f}  \ga^\mu D_\mu f,
\label{m4eq5a8}
\ee
one gets the interactions of the charged gauge bosons with leptons in the following Lagrangian part:
%----
\bea L_{\mathrm{leptons}} & = & \fr{g}{\sqrt{2}}\left[ \overline{\nu_{a L}} \ga^\mu \left(  W^{'+}_\mu l_{a L}  + Y_{ \mu}^{'-} l^c_{a L}+ N^0_{ \mu} \nu^c_{a L}\right)\right.\crn
&+&\left.  \overline{l_{a L}} \ga^\mu \left(  U_\mu^{--} l^c_{a L} + X_{ \mu}^{'-} \nu^c_{a L}\right)
+ \overline{l^c_{a L}} \ga^\mu K_\mu^{'+} \nu^c_{a L} \right] + \mathrm{H.c.}\crn
& = & \fr{g}{\sqrt{2}}\left[\frac{}{}  \overline{\nu_{a L}} \ga^\mu \left(  W'^+_\mu l_{a L}  + N^0_{ \mu} \nu^c_{a L}\right) - \overline{l_{a R}} \ga^\mu Y_{ \mu}^{'-}\nu^c_{a R}   \right.\crn
&+&\left.  \overline{l_{a L}} \ga^\mu \left(  U_\mu^{--} l^c_{a L} + X_{ \mu}^{'-} \nu^c_{a L}\right)
- \overline{ \nu_{a R}} \ga^\mu K_\mu^{'+} l_{a R}  \right] + \mathrm{H.c.}\crn
&=&  \fr{g}{\sqrt{2}}\left[\frac{}{} \overline{\nu_{a L}} \gamma^{\mu} N^0_{ \mu} \nu^c_{a L} +   \overline{l_{a L}} \ga^\mu  U_\mu^{--} l^c_{a L}\right.\crn
&+&\left. \overline{\nu_{a}} \ga^\mu \left( c_{\theta}P_L+ s_{\theta}P_R \right) l_{a}  W^+_\mu  +\overline{\nu_{a}} \ga^\mu \left( s_{\theta}P_L- c_{\theta}P_R \right) l_{a}  K^+_\mu \right. \crn
&+&\left.   \overline{l_{a}} \ga^\mu \left(  c_{\theta'}P_L - s_{\theta'}P_R \right) \nu^c_{a}X_{ \mu}^{-} - \overline{l_{a}} \ga^\mu \left(  s_{\theta'}P_L + c_{\theta'}P_R \right) \nu^c_{a }Y_{ \mu}^{-}  \right]+ \mathrm{H.c.},
\label{m4eq54b8}
\eea
where we have used
\[  \overline{l^c_{a L}} \ga^\mu  \nu^c_{a L} = - \overline{ \nu_{a R}} \ga^\mu l_{a R}\, .\]
From (\ref{m4eq54b8}), we see that  new  gauge bosons $K^\pm$ play a similar role of the SM $W^\pm$ for
 right-handed leptons with just opposite sign of coupling constant $g$.  This  right-handed current  also  appeared  in \cite{rhc}.
The gauge bosons carrying lepton number 2 (called bilepton gauge bosons) include: $Y^\pm, N^0, U^{\pm \pm}$, and
$X^\pm$.

For quarks, we have
\bea L_{\mathrm{quarks}} & = & \fr{g}{\sqrt{2}}\left[ \overline{u_{3 L}} \ga^\mu
\left(  W^{'+}_\mu d_{3 L}  + Y_{ \mu}^{'-} T_{ L}+ N^0_{ \mu} T^\prime_{ L}\right)
\right.\crn
&+&  \overline{d_{3 L}} \ga^\mu \left(  U_{ \mu}^{--} T_L + X_{ \mu}^{'-} T_L^\prime\right)
 + \overline{T_L} \ga^\mu K_{ \mu}^{'+} T_L^\prime \crn
 %-------den day 23/4
 &+ & \overline{d_{\al L} } \ga^\mu \left(  - W^{'-}_\mu u_{\al L}  + Y_{ \mu}^{'+} D_{\al L} + N_{ \mu}^{0 *}
 D_{\al L}^\prime\right)\crn
&-&\left.   \overline{u_{\al L}} \ga^\mu \left(  U_{ \mu}^{++} D_{\al L} + X_{ \mu}^{'+} D_{\al L}^\prime\right) + \overline{D_{\al L}} \ga^\mu K_{ \mu}^{'-} D_{\al L}^\prime \right] +\mathrm{  H.c.}.
\label{m4eq54c8}
\eea
Taking into account  the mixing among singly charged gauge bosons, we can express the above expression as
follows,
\bea - \mathcal{L}^{\mathrm{CC}}=\fr{g}{\sqrt{2}}\left(J^{\mu-}_W W^+_\mu + J^{\mu-}_K K^+_\mu + J^{\mu -}_X X^{+}_\mu +
J^{\mu-}_Y Y^+_\mu + J^{\mu 0*}_N N^{0}_\mu +  J^{\mu --}_U U^{++}_\mu +\mathrm{H.c.}\right), \eea
where
\bea J^{\mu-}_W&=&c_\theta (\overline{\nu}_{aL}\ga^\mu
l_{aL}+ \overline{u_{3 L}} \ga^\mu d_{3 L}- \overline{u}_{\al L}\ga^\mu d_{\al L})
\crn &-& s_\theta
(- \overline{ \nu_{a R}} \ga^\mu  l_{a R}   + \overline{T_L} \ga^\mu  T_L^\prime +
 \overline{D^\prime_{\al L}} \ga^\mu D_{\al L}),  \label{lm61}\\
 J^{\mu-}_K&=&c_\theta (- \overline{ \nu_{a R}} \ga^\mu  l_{a R}   + \overline{T_L} \ga^\mu  T_L^\prime +
 \overline{D^\prime_{\al L}} \ga^\mu D_{\al L}) + s_\theta
 (\overline{\nu}_{aL}\ga^\mu
l_{aL}+ \overline{u_{3 L}} \ga^\mu d_{3 L} - \overline{u}_{\al L}\ga^\mu d_{\al L}),\crn
J^{\mu -}_X &=& c_{\theta'}(\overline{\nu^c_{a L}} \ga^\mu l_{a L}  + \overline{ T'_L} \ga^\mu d_{3 L}
- \overline{u_{\al L}} \ga^\mu D_{\al L}^\prime) +
s_{\theta'}( \overline{l^c_{a L}} \ga^\mu  \nu_{a L} + \overline{T_{ L}} \ga^\mu u_{3 L} + \overline{d_{\al L} } \ga^\mu  D_{\al L})\, , \crn
J^{\mu -}_Y &=& c_{\theta'}
( \overline{l^c_{a L}} \ga^\mu  \nu_{a L} + \overline{T_{ L}} \ga^\mu u_{3 L} + \overline{d_{\al L} } \ga^\mu  D_{\al L})
- s_{\theta'}
(\overline{\nu^c_{a L}} \ga^\mu l_{a L}  + \overline{ T'_L} \ga^\mu d_{3 L}
- \overline{u_{\al L}} \ga^\mu D_{\al L}^\prime)\, ,
 \crn
J^{\mu --}_U & = & \overline{l^c_{a L}} \ga^\mu  l_{a L}  + \overline{T_L} \ga^\mu d_{3 L}
  - \overline{u_{\al L}} \ga^\mu  D_{\al L}\, ,
\crn
J^{\mu 0*}_N & = & \overline{\nu_{a L}} \ga^\mu \nu^c_{a L} +
\overline{u_{3 L}} \ga^\mu T^\prime_{ L} +
\overline{D_{\al L}^\prime } \ga^\mu
 d_{\al L}.
\label{lmt} \eea
It is emphasized that in the above expression, all fermions are in the weak states. For precision,
they should be in the mass states. For the latter case,
in the quark sector, the CKM matrix will  appear.
In the model under consideration, due to the neutrino  Dirac mass matrix, the lepton mixing
 matrix $V_{PMNS}$  will  appear  in the $J^{\mu-}_W$.
So in terms of mass eigenstates, the current in (\ref{lm61})
has a new  form:
\bea
J^{\mu-}_W&=&c_\theta (\overline{\nu}_{iL}\ga^\mu V^{i j}_{PMNS}
l_{jL}+  s_\theta
 \overline{ \nu_{i R}} \ga^\mu V^{i j}_{PMNS} l_{j R}) + \cdots.
\label{lm61t}
\eea
The neutral currents, including the electromagnetic current, are
\bea - \mathcal{L}^{\mathrm{NC}}=  e  J^\mu_{em} A_\mu  + \fr{g_4}{2 c_W}\sum_{i=1}^3 Z_\mu^i
\sum_f \{ {\bar f} \ga^\mu[g^{(V)}(f)_{i V} - g^{(A)}(f)_{i A} \ga_5] f\},
\label{may221}\eea
where \be  e = g \sin \theta_W\, , \, t= \fr{g^\prime}{g} =\fr{2 \sqrt{2} \sin \theta_W}{\sqrt{1-4 \sin^2 \theta_W}},\label{may222}\ee
and $Z^{1,2,3}$, which can be identified as $Z^1\simeq Z$ and $Z^{2,3}\simeq Z'_{3,4}$, are exact eigenstates of the matrix (\ref{m2NGRHNp}).

The neutral currents are similar to that shown in Ref. \cite{pp}, so  the reader is referred to the mentioned work.
Similar to the 3-3-1 models \cite{fcnc331}, in the model under consideration, there are FCNCs at the tree level due to $Z^2$ and $Z^3$.

The formula (\ref{may222}) leads to a consequence,
\be  \sin^2 \theta_W < 0.25  \label{may223}, \ee
which is the same as in the minimal 3-3-1 model.
As mentioned in (\ref{ftt}), this constraint is the same for the original
version \cite{flt,pp} and for the version we are considering now.

%------sua dot 1
With the   particle content in both the fermion and Higgs sectors similar to that in \cite{fa1}, we suggest
that the unification mass is in the range of ${\cal O}(10)$  TeV.  The possible Landau pole similar to those
in the minimal 3-3-1 model  \cite{landau331} will be considered in our future work. We note that some
interesting aspects relating to the Landau poles of  both the minimal 3-3-1 models and 3-4-1 models were indicated in \cite{ann}.
%--------het
%
\section{\label{potential}Higgs potential}
The most general potential can then be written in the
following form:
 \[ V(\eta,\rho,\ph, \chi, H) = V(\eta,\rho,\ph,\chi) + V(H),\]
where
\bea
V(\eta,\rho,\ph,\chi)&=& \mu^2_1 \eta^\dag \eta +
 \mu^2_2  \rho^\dag \rho +  \mu^2_3  \ph^\dag \ph + \mu^2_4  \chi^\dag \chi \crn
 &+&\lambda_1 (\eta^\dag \eta)^2 + \lambda_2(\rho^\dag \rho)^2 +
\lambda_3  (\ph^\dag \ph)^2  + \la_4 ( \chi^\dag \chi)^2 \crn
& + & (\eta^\dag \eta) [ \lambda_5 (\rho^\dag \rho) +
\lambda_6 (\ph^\dag \ph)  + \lambda_7 (\chi^\dag \chi) ]\crn
&+& \ (\rho^\dag \rho)[ \lambda_8(\ph^\dag \ph) + \la_9 (\chi^\dag \chi)] +\lambda'_9 (\ph^\dag \ph) (\chi^\dag \chi) \crn
& + & \lambda_{10} (\rho^\dag \eta)(\eta^\dag \rho) +
\lambda_{11} (\rho^\dag \ph)(\ph^\dag \rho) + \lambda_{12} (\rho^\dag \chi)(\chi^\dag \rho) \crn
& + &  \lambda_{13} (\ph^\dag \eta)( \eta^\dag \ph) +\la_{14}\, (\chi^\dag \eta)( \eta^\dag \chi)
+\la_{15}\, (\chi^\dag \ph)( \ph^\dag \chi) \crn
&+&(f \ep^{i j k l} \eta_i \rho_j \ph_k \chi_l +\mathrm{ H.c.}),
\label{m4eq5461}
\eea
and $V(H)$ consists of the  lepton-number-conserving (LNC) and -violating (LNV) parts, namely,
%--
\bea V(H)
 & \equiv & V_{\mathrm{LNC}} + V_{\mathrm{LNV}}, \label{may61} \\
 V_{\mathrm{LNC}} & = &
   \mu^2_5 \mathrm{Tr }(H^\dag H)  + [ f_{4} \chi^\dagger  H\eta^* + \mathrm{H.c.}]
+ \la_{16} \mathrm{Tr}[(H^\dag H)^2] +
\la_{17} [\mathrm{Tr}(H^\dag H)]^2
 \crn &&+ \mathrm{Tr}(H^\dag H)[\la_{18} (\eta^\dag\eta) +
 \la_{19} (\rho^\dag\rho) + \la_{20} (\ph^\dag\ph) + \la_{21} (\chi^\dag\chi) ]
\crn &&+\la_{22}(\chi^\dagger H)(H^\dag\chi)  + \la_{23}(\eta^\dag H)(H^\dag\eta)
+\la_{24}(\rho^\dagger H)(H^\dag \rho) +\la_{25}(\ph^\dagger H)(H^\dag \ph )
,\crn
%--------------
V_{\mathrm{LNV}} &=&  f_{2}
\chi^\dagger  H\chi^* + f_{3}  \eta^\dagger H  \eta^*  +  \la_{26} \mathrm{Tr}(H^\dag H)(\chi^\dag\eta) + \la_{27}(\chi^\dag H)(H^\dag\eta) + \mathrm{H. c.} .
 \label{may62}
\eea

In the below illustration for the Higgs spectrum we consider only the LNC part of $V(H)$.  The minimum conditions correspond to the six linear coefficients of the neutral Higgs bosons with non-zero VEVs that vanish, leading to the six equalities shown in  Appendix \ref{mcond}.

The squared mass matrix $\mathcal{M}^2_{\mathrm{DCH}}$ of the doubly charged Higgs (DCH) bosons is   shown in Appendix \ref{mcond}.  It can be checked that det $\mathcal{M}^2_{\mathrm{DCH}}$=0, so that there exist two Goldstone bosons of the $U^{\pm\pm}$ bosons. They can be found exactly as  follows:
\be G^{\pm\pm}_U= \frac{\sqrt{2}v' H^{\pm\pm}_1-\sqrt{2}v' H^{\pm\pm}_2-v\rho^{\pm\pm} +w\phi^{\pm\pm}}{\sqrt{w^2+v^2+4v'^2}}. \label{Gpmpm} \ee
There are three  physical masses. In the limit $v'^2\simeq0$, these masses are
\bea  m^2_{h^{\pm\pm}_1}&=&\frac{1}{4}(-\la_{25}w^2+\la_{24}v^2)=-m^2_{h^{\pm\pm}_2}, \hs   m^2_{h^{\pm\pm}_3} =\frac{w^2+v^2}{2} \left(\la_{11}-\frac{fVu}{wv}\right),\label{DCHmass0}\eea
where $h^{\pm\pm}_i$, $i=1,2,3$ are mass eigenstates of the DCHs. Hence, in the limit $v'=0$, there always exists a negative value of $-|\frac{1}{4}(-\la_{25}w^2+\la_{24}v^2)|$, implying a negative squared mass at the tree level. On the other hand, when $v'\neq0$, the matrix $\mathcal{M}^2_{\mathrm{DCH}}$ in (\ref{DCH}) gives a tree-level mass relation, Tr$(\mathcal{M}^2_{\mathrm{DCH}})=\sum_{i=1}^3m^2_{h^{\pm\pm}_i}$, or equivalently,
\be m^2_{h^{\pm\pm}_1}+m^2_{h^{\pm\pm}_2}+m^2_{h^{\pm\pm}_3}=\la_{16}v'^2+\frac{w^2+v^2}{2}\left(\la_{11}-\frac{fVu}{wv}\right). \label{DCHmre} \ee
As a consequence of the vacuum stabilities that the Higgs potential must be bounded from below, we have $\la_{16}>0$. Then the sum of the two squared DCH masses (\ref{DCHmre}) is  in the order of $\mathcal{O}(\la_{16}v'^2)$.  Because the DCH are solutions of the equation det$\left(\mathcal{M}^2_{\mathrm{DCH}}-I_4\times m^2_{h^{\pm\pm}}\right)=0$, we have another relation:
\be m^2_{h^{\pm\pm}_1}m^2_{h^{\pm\pm}_2}m^2_{h^{\pm\pm}_3}=-\frac{1}{16}(-\la_{25}w^2+\la_{24}v^2)^2\times \frac{w^2+v^2+4v'^2}{2}\left(\la_{11}-\frac{fVu}{wv}\right). \label{proDCH}\ee
The right-hand side of (\ref{proDCH}) is nonpositive because the factor $\left(\la_{11}-\frac{fVu}{wv}\right)$ has the same positive sign with squared masses of the heavy DCH $h^{\pm\pm}_3$. Hence, there is always a negative squared mass of DCH at the tree level. To avoid DCH tachyons, the $|\la_{25}w^2-\la_{24}v^2|$ should be small so that the loop contributions can raise the DCH mass to  positive values. As a result, the parameter $|\la_{25}|$ should be very small, and the model predicts the existence of rather light DCHs.

There are 12 pairs of singly charged  Higgs  (SCH) components in the original basis. In the mass basis, there are four massless pairs which are Goldstone bosons of $W,\; X,\; Y,\;$ and $K$  gauge bosons.  In the limit $\ep=0$, the squared mass matrix in the original basis decomposes into four independent $3\times3$ matrices;  see the details in the Appendix. Each of them  has only one zero eigenvalue, implying the massless state, and two other  massive values. All of the massless states are
 \bea G^{\pm}_{1}&=&\frac{\left(-v' H^{\pm}_1-w\phi^{\pm}_1+u \eta^{\pm}_3 \right)}{\sqrt{w^2+u^2+v'^2}},\;
 G^{\pm}_{2}=\frac{\left(-V \chi^{\pm}_2+ v'H^{\pm}_4+v \rho^{\pm}_2 \right)}{\sqrt{V^2+v^2+v'^2}},\crn
 G^{\pm}_{3}&=&\frac{\left(w \phi^{\pm}_2-V \chi^{\pm}_3+v' H^{\pm}_3 \right)}{\sqrt{V^2+w^2+v'^2}},\;
 G^{\pm}_{4}=\frac{\left(-u \eta^{\pm}_2+ v \rho^{\pm}_1+v' H^{\pm}_2 \right)}{\sqrt{u^2+v^2+v'^2}}.
   \label{scGold}\eea
In this limit, it is easily to identify that $G^{\pm}_Y\equiv G^{\pm}_1$, $G^{\pm}_X\equiv G^{\pm}_2$, $G^{\pm}_K\equiv G^{\pm}_3$,  and  $G^{\pm}_W\equiv G^{\pm}_4$ which are respective  Goldstone bosons absorbed by $Y^{\pm}$, $X^{\pm}$, $K^{\pm}$, and $W^{\pm}$ bosons.

 With $v'\neq0$, the masses as well as mass eigenstates are a bit complicated. For illustration, it is enough to consider here the limit $v',\ep\rightarrow 0$.  The mass eigenvalues of the other eight pairs of SCHs are
\bea
 m^2_{h^\pm_1}&=&\frac{1}{4}(\la_{23}u^2-\la_{25}w^2),\hs
  m^2_{h^\pm_2}=\frac{1}{4}\left(\la_{23}u^2-\la_{24}v^2\right),  \crn
  m^2_{h^\pm_3}&=&\frac{u^2+v^2}{2}(\la_{10}- f\frac{wV}{uv}),\;
  m^2_{h^\pm_4}=   \frac{u^2+\omega ^2}{2}\left(\la_{13}- \frac{f vV}{wu}\right),\crn
   m^2_{h^\pm_5}&=&  \frac{V^2+\omega ^2}{2}\left(\la_{15}- \frac{f vu}{Vw}\right),
  m^2_{h^\pm_6}=  \frac{V^2+v ^2}{2}\left(\la_{12}- \frac{f wu}{Vv}\right),\crn
  m^2_{h^\pm_7}&=& \frac{1}{4}\left( \la_{22}V^2-\la_{25}w^2\right),\;
  m^2_{h^\pm_8}= \frac{1}{4}\left( \la_{22}V^2-\la_{24}v^2\right)
\label{SCH}\eea
%--
with the respective mass states as follows:
\bea h^\pm_{1}&\equiv& H^{\pm}_1, \hs   h^\pm_{2}= H^{\pm}_2 ,\hs
h^\pm_{3}= \frac{v\eta^{\pm}_2+ u\rho^{\pm}_1}{\sqrt{u^2+v^2}},\hs   h^\pm_{4}= \frac{u\phi^{\pm}_1+w\eta^{\pm}_3}{\sqrt{u^2+w^2}}, \crn
 h^\pm_{5}&=& \frac{V\phi^{\pm}_2+w\chi^{\pm}_3}{\sqrt{V^2+w^2}} ,\hs   h^\pm_{6}= \frac{v\chi^{\pm}_2+ V\rho^{\pm}_2}{\sqrt{V^2+v^2}},\hs
h^\pm_{7}\equiv  H^{\pm}_3, \hs  h^\pm_{8}\equiv H^{\pm}_4 . \nn \eea
%--
The model predicts two rather light SCHs, $ h^{\pm}_1$ and $h^{\pm}_2$, because $\lambda_{ij}$ should be in the order of $\mathcal{O}(1)$; $|\lambda_{25}w|$ is not too large; and  $u$,  $v'$ and $v$ are in the electroweak scale.

There are ten neutral Higgs components in the original basis.  Four of the ten CP-odd neutral Higgses are massless, of which the four independent combinations of them are four Goldstone bosons of the $Z,\;Z^2\; Z^3$ and  $N^0$ gauge bosons. But there is  still one more massless  state, which is exactly $H^0_3$,  at the tree level. In the limit $\ep\rightarrow0$,  the mass eigenstates of the CP-odd neutral Higgs are shown explicitly in Appendix \ref{mcond}.  There are five massive states, with eigenstates denoted as $H_{A_i}$ ($i=1,5$), where three imagined parts of $H^0_{1,2,4}$ are approximate egeinstates.  The condition of the positive mass of the  CP-odd neutral Higgs $H_{A_5}$ shows that $f<0$. There also exists one light CP-odd neutral Higgs boson.

  In the neutral sector, one of the ten CP-even neutral Higgs bosons is the Goldstone  boson of the $N^{0*}$ boson. The squared mass matrix separates into two submatrices, namely the $4\times4$ and $6\times6$ matrices. They are denoted as $\mathcal{M}^2_{1H^0}$ and $\mathcal{M}^2_{2H^0}$, corresponding to the two respective sub-bases $(\mathrm{Re}[H^0_1],\;\mathrm{Re}[H^0_4],\;\mathrm{Re}[\chi^0_1],\; \mathrm{Re}[\eta^0_4])^T$ and $(\mathrm{Re}[H^0_3],\;\mathrm{Re}[\chi^0_4],\;\mathrm{Re}[\phi^0_3],\;\mathrm{Re}[\rho^0_2],\; \mathrm{Re}[\eta^0_1],\;\mathrm{Re}[H^0_2])^T$.  The massless values are  contained in $\mathcal{M}^2_{1H^0}$. In the limit $\ep\rightarrow0$,  three other mass values are
  \bea m^2_{h^0_1} &=&\frac{1}{4}\left( 2\la_{23}u^2-\la_{24}v^2-2\la_{16}v'^2-\la_{25}w^2\right),\;  m^2_{h^0_2}=  \frac{V^2+u ^2}{2}\left(\la_{14}- \frac{f wv}{Vu}\right),\crn
   m^2_{h^0_3} &=&\frac{1}{4}\left( 2\la_{22}V^2-\la_{24}v^2-2\la_{16}v'^2-\la_{25}w^2\right).\label{mmtrix1}\eea
   The mass eigenstates $(h^0_1,\;h^0_3,\;h^0_3)$ and the Goldstone bosons $G_{N^{0*}}$ in this case are
   \bea  h^0_1&\equiv& \mathrm{Re}[H^0_1], \hs  h^0_3 \equiv \mathrm{Re}[H^0_4], \hs
   \left(
     \begin{array}{c}
       G_{N^{0*}} \\
       h^0_2 \\
     \end{array}
   \right)=
   \left(
     \begin{array}{cc}
      -\frac{V}{\sqrt{V^2+u^2}} & \frac{u}{\sqrt{V^2+u^2}} \\
     \frac{u}{\sqrt{V^2+u^2}} & \frac{V}{\sqrt{V^2+u^2}}\\
     \end{array}
   \right)
   \left(
     \begin{array}{c}
       \mathrm{Re}[\chi^0_1] \\
       \mathrm{Re}[\eta^0_4] \\
     \end{array}
   \right).
    \nn  \eea
   %--
 It can be seen that while the two last  are very heavy with the order of the $SU(3)_L$ and $SU(4)_L$ breaking scales, the first Higgs boson may be
 lighter because $|\la_{25}w^2|$ should not be large as discussed above. So it may be the SM  Higgs boson or
 that in the 750 GeV diphoton excess.

  Now consider the second mass matrix $\mathcal{M}^2_{2H^0}$.  From our investigation,  in general it is easy to check that det$\mathcal{M}^2_{2H^0}\neq0$. But if $v'=0$, $\mathcal{M}^2_{2H^0}$ has a massless value. In addition, if $v'=v=u=0$, the matrix has two massless values, implying that there may be two light CP-even 
neutral Higgs bosons. Hence one of them can be identified with the SM Higgs boson, meaning that the Higgs sector of 
the model under consideration is reliable. The main contributions to the four heavy Higgs bosons are
  \bea m^2_{h^0_4} &=&-f wV,\hs  m^2_{h^0_5} =  \frac{1}{4}\left( \la_{22} V^2-\la_{25} w^2\right),\crn
  m^2_{h^0_{6,7}} &=& \la_3w^2+\la_4 V^2 \pm \sqrt{(-\la_3w^2+\la_4 V^2)^2+\la '^2_9w^2V^2}.\label{HeaH2}\eea
  %--
  To conclude, in the Higgs sector, we would like to emphasize two important results. First, the above investigation can be applied for  the models
   where the  $10_S$ $H$ is not included.  Second, the model predicts many Higgs bosons with masses near the TeV range
    that today colliders can detect.  Hence this aspect of the  Higgs sector should be explored in more detail.
\section{\label{phen}Phenomenology}
Our aim in this section is to find some constraints on the parameters
of the model. From mixing of the singly charged gauge bosons, we have
some special features related to the SM $W$ boson.
\ben
\item  In the  model under sonsideration,  the $W$ boson has the
following  normal main decay modes: \bea W^-  & \rightarrow &
l\  \tilde{\nu}_l\  (l = e,\mu,\tau),\crn
 &  \searrow & u^c
d, u^c s, u^c b,  (u\rightarrow c),\label{wdn}\eea
 which are the same as
in the SM. Due to the $W - K$ mixing, we have other   modes related to right-handed
lepton counterparts, namely,
  \be W^- \rightarrow l_R  \tilde{\nu}_{l R } \ (l
= e,\mu,\tau). \label{wmd}\ee
It is easy to compute the tree-level decay widths as follows \cite{bardin}.
 The predicted total width for the $W$
decay into fermions is \bea \Ga^{\mathrm{tot}}_W=1.04\fr{\al
M_W}{2s^2_W}(1-s^2_{\theta})+\fr{\al M_W}{4s^2_W}.\eea
This is quite similar to the case of the economical 3-3-1 model \cite{eco}.
 From the recent data of the $W^{\pm}$ boson  \cite{data}: $\al(m_Z) \simeq 1/128$, $m_W=80.385\pm 0.015$ GeV,
 $\Ga^{\mathrm{tot}}_W=2.085\pm 0.042\, $ GeV.
The  $s_{\theta}$ is less constrained than that of  \cite{eco}, with an upper bound of $s_{\theta}\leq 0.19$.

\begin{figure}[h]
 \bc
\begin{picture}(100,100)(50,0)
\ArrowLine(40,50)(75,20)
 \ArrowLine(10,50)(40,50)
%-----
 \Photon(40,50)(70,50){2}{5}
 \ArrowLine(100,30)(70,50)
 \ArrowLine(70,50)(100,70) \Text(70,10)[]{$\nu_\mu$}
\Text(10,60)[]{$\mu^-$} \Text(58,60)[]{$W,K$}
 \Text(110,20)[]{$\tilde{\nu}_e$}
 \Text(110,80)[]{$e^-$}
\end{picture}
%--
\begin{picture}(100,100)(-30,0)
\ArrowLine(40,50)(75,20)
 \ArrowLine(10,50)(40,50)
 %--
 \Photon(40,50)(70,50){2}{5}
 \ArrowLine(100,30)(70,50)
 \ArrowLine(70,50)(100,70) \Text(70,10)[]{$\nu^c_\mu$}
\Text(10,60)[]{$\mu^-$} \Text(58,60)[]{$X,Y$}
 \Text(110,20)[]{$\nu^c_e$}
 \Text(110,80)[]{$e^-$}

\end{picture}
\ec \caption[]{ Feynman diagram giving contribution to muon decay. The left and right diagrams present the main and wrong decay channels, respectively. } \label{muondecay}
\end{figure}
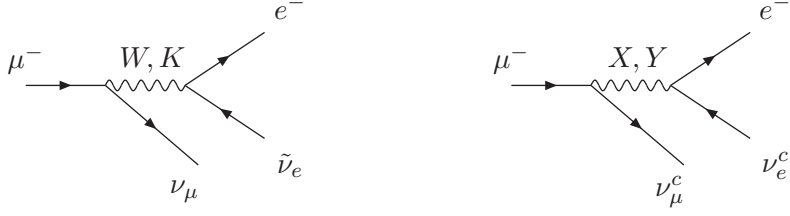

\item In the model under consideration, the muon  decay,
\[ \mu^-  \rightarrow e^- + \tilde{\nu}_e + \nu_\mu \],
 consists of two diagrams mediated by $W$ and $K$. The Feynman diagrams are  on the left in  Fig. \ref{muondecay}.
%-------

The decay width is given by
\be \Gamma(\mu^-  \rightarrow e^- + \tilde{\nu}_e + \nu_\mu)=\frac{g^4 m_\mu^5}{6144\pi^2}\times \left(\frac{1}{m_W^4}+\frac{1}{m_K^4}\right). \label{Gmu341dec}\ee
Because the model predicts the wrong decay $\mu^-  \rightarrow e^- + \nu_e + \tilde{\nu}_\mu$, we assume that the total decay of the muon is $ \Gamma^{\mathrm{total}}_{\mu}=  \Gamma(\mu^-  \rightarrow e^- + \, \tilde{\nu}_e +\,  \nu_\mu)+ \Gamma(\mu^-  \rightarrow e^-
+\,  \tilde{\nu}_\mu+ \, \nu_e )$.
This result will be used for in a future study.
%--
\item The wrong decay of the muon is
\[ \mu^-  \rightarrow e^- + \nu_e + \tilde{\nu}_\mu, \]
%--
where the Feynman diagram is on the right side of  Fig. \ref{muondecay}. The branching ratio of the wrong decay
Br$(\mu\rightarrow e\,  \nu_e  \,  \bar{\nu}_\mu ) < 0.012$ leads to the following constraint:
\be \frac{\frac{1}{m_X^4}+\frac{1}{m_Y^4}}{\frac{1}{m_W^4}+\frac{1}{m_K^4}+\frac{1}{m_X^4}+\frac{1}{m_Y^4}}<0.012 \hs \mathrm{or}\;  \frac{1}{m_X^4}+\frac{1}{m_Y^4} -0.012\left(\frac{1}{m_K^4} +\frac{1}{m_X^4}+\frac{1}{m_Y^4}\right) < \frac{0.012}{m_W^4}\, .  \label{wdecayconstrain}\ee
In the limit $V\gg w$,  i.e.,  $1/m^4_K,1/m^4_X\ll 1/m^4_Y$, we obtain the below constraint of $m_Y$: $m_Y > m_W\times \sqrt[4]{82.333}\simeq 242$ GeV. This is consistent with those in Ref. \cite{wmud}.  In the limit of $V\simeq w$, implying that $m_X^2 \simeq m_Y^2\simeq  m_K^2/2$, the constraint is  more strict:   $m_Y > m_W\times \sqrt[4]{164.417}\simeq  287$ GeV.

\item The $\mu-e$ conversion:
The charged current of  the  model under consideration  has a similar structure as the ones discussed in \cite{wmud,muconv1,muconv2,muconv3} so these have the similar results as the $\mu-e$ conversion.  In particular, the mass of the doubly charged bileptons $U^{\pm\pm}$ satisfies $m_{U^{\pm\pm}}\geq 135$ GeV \cite{muconv2}. Note that this result was concluded for the $SU(3)_L$, but it is the same for the $SU(4)_L$ because they have the same gauge couplings.  This constraint is less strict than the constraint from the wrong muon decay, because $m_{Y^{\pm}}$ and $m_{U^{\pm\pm}}$ are  related with only the $SU(3)_L$ breaking scale $w$.
\een

\section{\label{conc} Conclusion}

In this paper, we have analyzed the 3-4-1 model with arbitrary electric charges of the extra leptons. The scalar and gauge boson sectors are presented in detail . The  mixing matrix, the eigenmasses, and the eigenstates of neutral gauge bosons are analyzed.  For future studies, we will focus on the scalar sector, especially neutral one, in which the SM-like Higgs boson is contained.

Next, we have presented a new development of the original  3-4-1 model.
  Different from previous  works \cite{flt,pp}, in this paper, with the assumption of a new nonzero VEV of a neutral Higgs component in  the decuplet $H$,  some new interesting features occur: (i) the neutrinos get  Dirac masses at tree level;  ii) the mixing among the singly charged gauge bosons leads to the appearance of a new small contribution to the decay width of the $W$ boson, the same as that shown in the economical 3-3-1 model. But under the present $W$ data, the constraint of the mixing angle is less strict. The model also predicts the existence of many bileptons, including new quarks, and gauge and Higgs bosons, as well as the LFV interactions. Like the 3-3-1 models, many of these bileptons contribute to the LFV processes such as the wrong muon decay, and the $\mu-e$ conversion.   If the $SU(4)_L$ breaking scale is much larger than the $SU(3)_L$ breaking scale, so that it gives suppressed  contributions to the LFV processes, the constraint of the $SU(3)_L$ breaking is the same as what was indicated in the 3-3-1 models. In contrast, if the two breaking scales are close together, the lower bounds of the $SU(3)_L$ breaking scale significantly  increase.  We see this point in the case of the  wrong muon decay, where the lower constraints of $m_Y$ are $242$ and $287$ GeV, corresponding to the two mentioned cases.

We have also derived the lepton number operator, and the lepton number of the fields in the model is  presented.

As in the minimal 3-3-1 model, in the 3-4-1 model with right-handed neutrino considered here,   there exists a bound on the sine-squared of the Weinberg angle, namely, $\sin^2 \theta_W < 0.25$. The constraint on the electric charges of extra leptons has been   obtained as well.

The Higgs sector is roughly studied. In the limit of lepton number conservation, the Higgs sector contains all massless Goldstone bosons for massive gauge bosons and the SM-like Higgs  boson.

The  model we have considered  is quite interesting  and deserves further study.
%-----
\section*{Acknowledgments}
  LTH thanks Le Duc Ninh for useful discussions on gauge anomalies and SM matching conditions. This research is funded by the Vietnam  National Foundation for Science and Technology Development (NAFOSTED) under grant number 103.01-2014.51.
\appendix
\section{\label{mcond} Higgs spectrum}
This section pays attention to the Higgs potential satisfying the lepton number conservation.  In addition,  the squared mass matrices are written in the limit of $\ep=0$,
 except  the squared mass matrix of the doubly charged Higgs which is independent of $\ep$.
\subsection{ Minimal conditions of the Higgs potential}
We list here six equalities  for minimal conditions of the Higgs potential:
{%\footnotesize
\bea \mu_1^2&=& - \frac{1}{4}\left[ 2\la_{16}\ep^2(v'^2-\ep^2) -\la_{22}\ep^2V^2+\la_{25}\ep^2w^2 +\la_{24}\ep^2v^2+2fVwvu \right]\crn
&-&\frac{1}{2}\left[\la_{18} v''^2+2\la_1u^2+\la_5v^2+\la_6 w^2+\la_7 V^2 \right], \crn
%-
 \mu_2^2&=& - \frac{1}{2}\left[ 2\la_2 v^2+\la_5 u^2+ \la_8 w^2+\la_9 V^2 + \frac{f w Vu}{v}+  \la_{19} v''^2 +\frac{1}{2}\la_{24}  v'^2 \right] ,\crn
 %-
  \mu_3^2&=& - \frac{1}{2}\left[ 2\la_3 w^2 +\la_6 u^2 + \la_8 v^2+ \la'_9 V^2 + \frac{fuVv}{w}+ \la_{20}v''^2+\frac{1}{2} \la_{25}v'^2 \right],\crn
  %-
   \mu_4^2&=& - \frac{1}{2}\left[2 \la_4 V^2+  \la_7 u^2 +\la_9 v^2 +\la'_9 w^2 +\la_{21}v''^2  \right]\crn
   &-& \frac{1}{4 V^2}\left[2 f Vwuv +2\la_{16}\ep^2(v'^2-\ep^2) -\la_{23}\ep^2u^2+ \la_{24}\ep^2 v^2 +\la_{25}\ep^2 w^2  \right],\crn
   \mu_5^2&=& - \frac{1}{2}\left[\la_{16}v'^2 +2\la_{17}v''^2 + \la_{18} u^2+ (\la_{19}+\frac{1}{2}\la_{24}) v^2 +\la_{20}w^2+\la_{21}V^2  +\frac{1}{2} \la_{25}w^2 \right],\crn
   f_4&=&  \frac{\ep}{2 Vu}\left[ 2\la_{16}(v'^2-\ep^2)- \la_{22} V^2 -\la_{23}u^2 +\la_{24} v^2+\la_{25} w^2 \right].
 \label{minimacond}\eea
}
%---
\subsection{Squared mass matrix of doubly charged Higgses}
Squared mass matrix of the doubly charged Higgses  in the basis $(H_1^{\pm\pm},\;H_2^{\pm\pm},\; \rho^{\pm\pm},\;\phi^{\pm\pm})^T$  is given by
%--
{\footnotesize
\bea  \mathcal{M}^2_{\mathrm{DCH}} =\frac{1}{4} \left(
                                       \begin{array}{cccc}
                                        2\la_{16}v'^2+\la_{24}v^2-\la_{25}w^2  & 2\la_{16}v'^2& \sqrt{2}\la_{24}vv' &  \sqrt{2}\la_{25}wv'\\
                                          &   2\la_{16}v'^2-\la_{24}v^2+ \la_{25}w^2 &  \sqrt{2}\la_{24}vv'  &  \sqrt{2}\la_{25}wv'   \\
                                          &  & 2\la_{11}w^2-\frac{2fwVu}{v} &  2\la_{11}w v -2fVu  \\
                                          &  &  & 2\la_{11} v^2-\frac{2fVvu}{w} \\
                                       \end{array}
                                     \right).\crn
 \label{DCH}\eea
 }
\subsection{Squared mass matrices of singly charged Higgses}
 The squared mass matrix of the singly charged Higgs consists of two independent $6\times 6$ matrices.  They are denoted as $\mathcal{M}^2_{1\mathrm{sch}}$ and $\mathcal{M}^2_{2 \mathrm{sch}}$ with respective sub-bases $(H_1^{\pm},\; \phi_1^{\pm} ,\; \eta_3^{\pm},\;\chi_2^{\pm},\;H_4^{\pm},\;\rho_2^{\pm} )^T$ and $(\phi_2^{\pm},\;\chi_3^{\pm},\;H_3^{\pm},\;\eta_2^{\pm},\;\rho_1^{\pm} ,\;H_2^{\pm})^T$. In the limit $\ep=0$, they divide into four independent $3\times3$ submatrices, denoted as
 \bea \mathcal{M}^2_{1\mathrm{sch}}=\left(
                                     \begin{array}{cc}
                                       \mathcal{M}^{\prime2}_{1\mathrm{sch}} & 0 \\
                                        0&   \mathcal{M}^{\prime\prime2}_{1\mathrm{sch}}\\
                                     \end{array}
                                   \right),\;\mathrm{ and} \;
  \mathcal{M}^2_{2\mathrm{sch}}=\left(
                                     \begin{array}{cc}
                                       \mathcal{M}^{\prime2}_{2\mathrm{sch}} & 0 \\
                                        0&   \mathcal{M}^{\prime\prime2}_{2\mathrm{sch}}\\
                                     \end{array}
                                   \right),\nn
  \label{}\eea
  where
   {\footnotesize
 \bea
 \left(\mathcal{M}^{\prime2}_{1\mathrm{sch}}\right) =\frac{1}{4}
 \left(
   \begin{array}{ccc}
   \la_{23}u^2-\la_{25}w^2    &\la_{25}wv'  &\la_{23}uv'  \\
      &2\la_{13}u^2-\la_{25}v'^2 -\frac{2fVvu}{w} &2(\la_{13}wu-fVv) \\
      &  & 2\la_{13}w^2+ \la_{23}v'^2 -\frac{2fVwv}{u}  \\
   \end{array}
 \right),
  \eea
 }
  %--
    {\footnotesize
 \bea
 \left(\mathcal{M}^{\prime\prime2}_{1\mathrm{sch}}\right) =
 \frac{1}{4} \left(
   \begin{array}{ccc}
     \la_{22}v'^2 +2\la_{12}v^2-\frac{2fwuv}{V}&\la_{22}Vv'  &2\left(\la_{12}Vv-fwu \right)  \\
      & \la_{22}V^2-\la_{24}v^2  &\la_{24}vv'  \\
      &  & 2\la_{12}V^2-\la_{24}v'^2 -\frac{2fVwu}{v} \\
   \end{array}
 \right),
  \eea
 }
 {\footnotesize
 \bea \left(\mathcal{M}^{\prime2}_{2\mathrm{sch}}\right) =
 \frac{1}{4}  \left(
  \begin{array}{ccc}
  2\la_{15}V^2 -\la_{25}v'^2-\frac{2fVvu}{w} &2\left(\la_{15}Vw -f vu \right) &\la_{25}v'w  \\
     &  2\la_{15}w^2 +\la_{22}v'^2-\frac{2fwvu}{V} &  \la_{22}Vv' \\
     &  & \la_{22}V^2- \la_{25}w^2 \\
   \end{array}
   \right),
  \eea
 }
 and
 {\footnotesize
 \bea
  \left(\mathcal{M}^{\prime\prime2}_{2\mathrm{sch}}\right) =
   \frac{1}{4}\left(
    \begin{array}{ccc}
   \la_{23}v'^2+2\la_{10}v^2-\frac{2 fVwv}{u}& 2\left( \la_{10}uv-fVw\right) &\la_{23}v'u \\
       &2\la_{10}u^2 -\la_{24}v'^2- \frac{2fVuw}{v} &\la_{24}vv' \\
       &  &  \la_{23}u^2-\la_{24}v^2\\
    \end{array}
  \right).
 \eea
 }
\subsection{Squared mass matrices of CP-odd neutral Higgses}
This $10\times 10$ matrix  has a massless state of $\mathrm{Im}[H^0_3]$, even $\ep\neq0$. Furthermore, the remaining part separates into two $4\times4$ and $5\times5$  matrices, corresponding to the bases of $(\mathrm{Im}[\chi^0_1],\; \mathrm{Im}[\eta^0_4],\; \mathrm{Im}[H^0_4],\; \mathrm{Im}[H^0_1])^T$ and $(\mathrm{Im}[\rho^0_2],\;\mathrm{Im}[\phi^0_3],\;\mathrm{Im}[\chi^0_4],\; \mathrm{Im}[\eta^0_1],\; \mathrm{Im}[H^0_2] )^T$. In the limit $\ep=0$, the following Higgses are identified with mass eigenstates:
\bea\mathrm{Im}[H^0_4] &\equiv& H_{A_1},\; m^2_{A_1}= \frac{1}{4}\left( 2\la_{22}V^2- 2\la_{16}v'^2 -\la_{24}v^2-\la_{25}w^2 \right), \crn
 \mathrm{Im}[H^0_1] &\equiv& H_{A_2},\; m^2_{A_2}=  \frac{1}{4}\left( 2\la_{23}u^2- 2\la_{16}v'^2 -\la_{24}v^2-\la_{25}w^2 \right), \crn
  \mathrm{Im}[H^0_2] &\equiv& H_{A_3},\; m^2_{A_3}=  \frac{1}{4}\left( \la_{22}V^2+\la_{23}u^2- 2\la_{16}v'^2 -\la_{24}v^2-\la_{25}w^2 \right). \label{cpoddH1} \eea
 The nontrivial parts of the two matrices are  now $2\times2$ and $4\times4$ which relate to the four Goldstone bosons, including $G_{N^0}$ and  $G_{Z_i}$ ($i=1,2,3$),  and the two massive CP-odd neutral Higgses, in particular
 {\footnotesize
  \bea \left(
        \begin{array}{c}
          G_{N_0} \\
           H_{A_4} \\
        \end{array}
      \right)= \left(
                 \begin{array}{cc}
                   \frac{V}{\sqrt{V^2+u^2}} &  \frac{u}{\sqrt{V^2+u^2}}  \\
                       -\frac{u}{\sqrt{V^2+u^2}}&    \frac{V}{\sqrt{V^2+u^2}} \\
                 \end{array}
               \right)
  \left(
        \begin{array}{c}
         \mathrm{Im}[\chi^0_1] \\
           \mathrm{Im}[\eta^0_4] \\
        \end{array}
      \right)
  \eea
 }
 and
 {\footnotesize
 \bea
 \left(
   \begin{array}{c}
     G_{Z_1} \\
     G_{Z_2} \\
     G_{Z_3} \\
   H_{A_5} \\
   \end{array}
 \right)=
 \left(
   \begin{array}{cccc}
    - \frac{v}{\sqrt{v^2+u^2}} & 0 & 0 & \frac{v}{\sqrt{v^2+u^2}} \\
     - \frac{vu^2}{\sqrt{A(v^2+u^2)}} &  0&   \frac{\sqrt{(v^2+u^2} V}{\sqrt{A}} &-\frac{uv^2}{\sqrt{A(v^2+u^2)}} \\
     - \frac{V^2u^2v}{\sqrt{AB}}  &  \frac{\sqrt{A}w}{\sqrt{B}} &  -\frac{Vv^2u^2}{\sqrt{AB}} & -\frac{V^2v^2u}{\sqrt{AB}}  \\
     \frac{Vwu}{\sqrt{B}}  &  \frac{Vvu}{\sqrt{B}} & \frac{wvu}{\sqrt{B}}  & \frac{Vwv}{\sqrt{B}}  \\
   \end{array}
 \right)
 \left(
   \begin{array}{c}
   \mathrm{Im}[\rho^0_2] \\
     \mathrm{Im}[\phi^0_3] \\
    \mathrm{Im}[\chi^0_4] \\
  \mathrm{Im}[\eta^0_1] \\
   \end{array}
 \right),
 \eea
 }
where $A=V^2v^2+u^2(V^2+v^2)$, $B=V^2v^2(w^2+u^2)+ w^2u^2(V^2+v^2)$.  Note that the Goldstone bosons of the three Hermitian gauge bosons are linear combinations of the three above massless states $G_{Z_i}$, not exactly themselves. But the $G_{Z_1}$ mainly contributes to the Goldstone boson of the SM $Z$ boson.  The masses of the $H_{A_{4,5}}$ are
 \be m^2_{A_4}= \frac{(V^2+u^2)}{2}\left(\la_{14}-\frac{fwv}{Vu}\right),\hs m^2_{A_5}= -\frac{f}{2}\left[ \frac{Vvu}{w}+w\left(\frac{Vv}{u}+\frac{u(V^2+v^2)}{Vv}\right)\right].\label{mha45}\ee
 %--

\end{document}